\documentclass[11pt, letter]{emulateapj}
\shorttitle{Dynamical Outcomes of Planet--Planet Scattering}
\shortauthors{Chatterjee, Ford, \& Rasio}

\begin{document}
\title{Dynamical Outcomes of Planet--Planet Scattering}
\author{Sourav Chatterjee,\altaffilmark{1} Eric B.\ Ford,\altaffilmark{2,3,4} Soko Matsumura,\altaffilmark{1} and Frederic A.\ Rasio\altaffilmark{1}}
\altaffiltext{1}{Department of Physics and Astronomy, Northwestern University, 
Evanston, IL 60208}
\altaffiltext{2}{Harvard-Smithsonian Center for Astrophysics, Mail Stop 51, 60 Garden Street, 
Cambridge, MA 02138}
\altaffiltext{3}{Department of Astronomy, University of Florida, 211 Bryant Space Science Center, P.O.\ Box 112055, Gainesville, FL, 32611}
\altaffiltext{4}{Hubble Fellow}

\begin{abstract}
Observations in the past decade have revealed extrasolar planets with
a wide range of orbital semimajor axes and eccentricities.  Based on the
present understanding of planet formation via core accretion and
oligarchic growth, we expect that giant planets often form in closely
packed configurations.  While the protoplanets are embedded in a
protoplanetary gas disk, dissipation can prevent eccentricity growth and
suppress instabilities from becoming manifest.  However, once the
disk dissipates, eccentricities can grow rapidly, leading to close
encounters between planets. Strong planet--planet
gravitational scattering could produce both high
eccentricities and, after tidal circularization, very short-period planets,
as observed in the exoplanet population.
We present new results for this scenario based on extensive dynamical 
integrations of systems containing three giant planets, both with and without residual
gas disks. We assign the initial planetary masses and orbits in a realistic manner
following the core accretion model of planet formation.  
 We show that, with realistic initial conditions, planet--planet 
scattering can reproduce quite well the observed eccentricity distribution.  
Our results also make testable predictions for the orbital {\em inclinations\/} of
short-period giant planets formed via strong planet
scattering followed by tidal circularization.
\end{abstract}


\section{Introduction}
\label{intro}

\begin{figure}
\begin{center}
\plotone{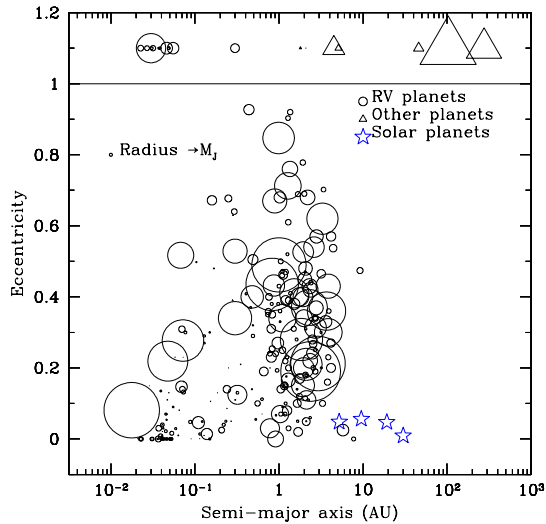}
\caption{Semi-major axis vs eccentricity for the detected
planets.  The sizes of the points are proportional to the minimum
masses ($m\sin i$) of the corresponding planets.  The size of a
Jupiter mass planet is shown at the left top corner for reference.
The stars represent the four giant planets in our solar system (for
these the sizes do not indicate their mass).  The open circles show the planets
detected by radial velocity surveys.  The triangles show planets
detected by micro-lensing or direct imaging.  Planets with poorly constrained
eccentricities are plotted above e = 1.  A horizontal line is drawn at
e = 1 to guide the eye.  Note the logarithmic scale for the semi-major
axis.}
\label{a_e_obs}
\end{center}
\end{figure}

The study of extrasolar planets and their properties has become a very
exciting area of research over the past decade.  Since the detection
of the planet 51~Peg~b, more than $200$ new planets \citep[][see also
http://exoplanet.eu/]{2006ApJ...646..505B} have been detected and the
large sky surveys planned for the near future can potentially detect
many more.  These detections have raised many questions about the
formation and dynamical evolution of planetary systems.  The
extrasolar planet population covers a much greater portion of the
semimajor axis and eccentricity plane than was expected based on the
planets in our solar system \citep[][Fig.~\ref{a_e_obs}]{1995Icar..114..217L}.  
The presence of many giant
planets in highly eccentric orbits or in very short-period orbits (the
``hot Jupiters'') is particularly puzzling.

  Different scenarios have been proposed to explain the high
eccentricities.  The presence of a distant companion in a highly
inclined orbit can increase the eccentricities of the planets around a
star through Kozai oscillations
\citep{1997ApJ...477L.103M,1997Natur.386..254H}.  However, this alone
cannot explain the observed eccentricity distribution
\citep{2005ApJ...627.1001T}.  Interaction with the protoplanetary gas
disk could either excite or damp the eccentricities depending on the
properties of the disk and the orbits.  However, the combined effects
typically result in eccentricity damping
\citep{1992PASP..104..769A,2001MNRAS.325..221P,2003ApJ...585.1024G,2003ApJ...587..398O}.
Migration of two planets and trapping in a mean-motion resonance (MMR)
can also pump up the eccentricities efficiently, but this mechanism
requires strong damping at the end or termination of migration right
after trapping in resonance \citep{2002ApJ...567..596L} or else it leads 
to planet scattering \citep{2006A&A...451L..31S}.
\citet{2004AJ....128..869Z} proposed inward propagation of
eccentricity after the outer planets are excited to high
eccentricities following a close encounter with a passing star.  Using
typical values for such interactions with field stars in the solar
neighborhood, however, they do not get very high eccentricities.
\citet{2001MNRAS.325..221P,2002MNRAS.332L..39T,1997ApJ...490L.171B}
propose a very different formation scenario for planets from
protostellar collapse in which both hot Jupiters and eccentric planets
at higher semimajor axes are formed naturally; this scenario,
however, cannot form sub-Jupiter-mass planets.

In this paper, we explore another promising way to create high
eccentricities: strong gravitational scattering between planets in a
multi-planet system undergoing dynamical instability
\citep{1996Sci...274..954R,1996Natur.384..619W,1997ApJ...477..781L}.
According to the model of oligarchic growth, planetesimals form in a
nearly maximally packed configuration in the protoplanetary disk, 
followed by gas accretion
\citep{2004ApJ...614..497G,2004ApJ...616..567I,2002ApJ...581..666K}.
Once the disk dissipates, mutual planetary perturbations (``viscous
stirring'') of the planetesimals will lead to eccentricity growth,
orbit crossing, and eventually close encounters between the big bodies
in the disk \citep{2007ApJ...661..602F, 2007Icar..189..196L}.  While
planetary systems with more than two planets can not be provably
stable, they can remain stable for very long timescales depending on
their initial separations
\citep{1996Icar..119..261C,2002Icar..156..570M}.  A sufficiently
massive disk can prevent interacting planets from acquiring large
eccentricities and developing crossing orbits.  However, once the gas disk
is sufficiently dissipated, and the planetesimal disk depleted, 
the eccentricities of the planets can grow
to high values, possibly leading to strong planet--planet scattering
and a phase of chaotic evolution that dramatically alters the orbital
structure of the system
\citep{1997ApJ...477..781L,2007ApJ...661..602F,2007Icar..189..196L}.

The detection of close-in planets with orbital periods as short as
$\sim 1\,$d, the so-called ``hot Jupiters'' (and, more recently, hot
Neptunes and super-Earths), was another major surprise.  Giant planets
are most likely to form at much larger separations, beyond the ice
line of the star where there can be enhanced dust production
\citep{2002ApJ...581..666K,2004ApJ...616..567I}.  It is widely
believed that the giant planets form beyond the ice line and then
migrate inwards to form the hot Jupiters we observe today.  Different
stopping mechanisms of inward migration have been proposed to explain the
hot Jupiters, but it is unclear why they pile up at just a few solar
radii around the star, rather than continue migrating and eventually
accrete onto the star.

  Strong gravitational scattering between planets in a multi-planet
system may provide another way to create these close-in planets
\citep{1996Sci...274..954R}.  A few of the planets scattered into very
highly eccentric orbits could have sufficiently small periastron
distances that tidal circularization takes place, giving rise to the
hot Jupiters.  The currently observed edge in the mass-period diagram
is very nearly at the ideal circularization radius (twice the Roche
limit), providing support for this model \citep{2006ApJ...638L..45F}.
\citet{2005Icar..175..248F} finds that these violent passages might
not destroy the planets, even if mass loss occurs.

 Previous studies have investigated gas free systems with two planets
around a central star extensively
\citep{1996Sci...274..954R,2001Icar..150..303F,
2003ASPC..294..181F,2007astro.ph..3163F} and have also begun to investigate
the dynamics of two planets in the presence of a gas disk by using
simplified prescriptions for dissipative effects
\citep{2005Icar..178..517M}.  The observed
eccentricity distribution is not easily reproduced by two equal-mass
planets \citep{2001Icar..150..303F}.  However,
strong scattering of two {\it unequal-mass} planets could explain the
observed eccentricities of most observed exoplanets
\citep{2003ASPC..294..181F,2007astro.ph..3163F}.  

A system with three planets is qualitatively different than
one with two planets. In two-planet systems, there is a sharp
boundary between initial conditions that are provably Hill stable
and initial conditions that quickly lead to a close encounter
\citep{1993Icar..106..247G}.  Moreover, for two Jupiter-mass planets
in close to circular orbits, this boundary lies where the ratio of orbital periods 
is less than 2:1.  
If these planets formed further apart,
then a slow and smooth migration could lead to systems becoming
trapped in a 2:1 MMR before triggering an instability 
(\citealt{2002ApJ...567..596L}; but see \citealt{2006A&A...451L..31S}).
These stability properties are in sharp contrast to those of systems
with three or more planets, for
which there is no sharp stability boundary. These systems can become
unstable even for much wider initial spacings and the timescale to the
first close encounter can be very long \citep{1996Icar..119..261C}.
Even if all pairs of adjacent planets in the system are stable according
to the two-planet criterion, the combined
system can evolve in a chaotic (but apparently bounded) manner for an
arbitrarily long time period before instability sets in.  This timescale to
instability can easily exceed other timescales of interest here such as 
those for the formation
of giant planets, orbital migration, or dispersal of the gas disk.
Therefore, the stability properties of three-planet systems can be studied with 
long-term orbital integrations without the need to implement any 
additional physics.

A pioneering study of the stability and final orbital properties of
three-planet systems was performed by
\citet[][hereafter MW02]{2002Icar..156..570M}.  They explored the basic
nature of instabilities arising in systems with three giant planets
around a central solar-like star, and they determined the final orbital
properties after one planet is ejected.  This study used highly
idealized initial conditions with three
equal-mass planets or with one arbitrary mass
distribution (middle planet twice as massive as the other two). It was 
also computationally limited to a rather small ensemble of systems.  

Our two main goals in this new work were to extend the work of MW02 by  
using more realistic initial conditions (see \S\ref{gas_free})
and to perform a much more extensive numerical survey in order to 
fully characterize the statistical properties of outcomes for unstable
three-planet systems.  
Given the chaotic nature of the
dynamics, one needs to integrate many independent systems to characterize the
statistical properties of the final planetary orbits
\citep[see a detailed analysis for the dependence of various statistics on the sample size in 
Appendix~\ref{size_statistic}; see also ][]{2003Icar..163..290A}.  Each system has to be integrated for a
long time, so that it reaches the orbit-crossing unstable phase and
later evolves into a new, stable configuration.  Given the rapid increase
in computing power, we are now able to perform significantly more and
longer integrations to obtain much better statistical results than was
possible just a few years ago.

In addition, we also present the results of new simulations for systems of three
giant planets still embedded in a residual gas disk. Here our goal is different:
we focus on the transition from gas-dominated to gas-free systems, in the hope of 
better justifying the (gas-free) initial conditions adopted in the first part of our work.
However, implementing the physics of planet-disk interactions implies a considerably 
higher computational cost, preventing us from doing a complete statistical study of 
outcomes at this point.

\section{Gas-free Systems with Three Giant Planets}
\label{gas_free}

In this section we consider systems with
three unequal-mass giant planets orbiting a central star of mass $1\,M_{\odot}$ 
at distances of several AU. The planets interact with each other through gravity 
and physical collisions only. We first present our assumptions and initial conditions,
with particular emphasis on realistic mass distributions for the planets, and then
we describe our numerical results and their implications.

\subsection{Initial Orbits}
\label{ini_orbits}

For all systems the initial
semimajor axis of the closest planet is always set at $a_1 = 3\,{\rm AU}$. The other two
planets are placed using the spacing law introduced by MW02,
\begin{equation}
a_{i+1} = a_i + K R_{H,i},
\label{spacing_planet}
\end{equation} 
where $R_{H,i}$ is the Hill radius of the $i^{th}$ planet and
we set $K = 4.4$ for all runs in this section.
These choices are somewhat arbitrary, but are
guided by the following considerations. For a solar-mass central star,
 the ice line is around $3\,\rm{AU}$ \citep{2002ApJ...581..666K}
and it is difficult to form giant planets closer to the star \citep[see e.g.,][]{2002ApJ...581..666K}.   
Although inward type~II migration \citep[see e.g.,][]{1980ApJ...241..425G} 
can bring the giant planets closer 
to the star, we avoid putting the planets initially very close to the star since 
very small initial semi-major axis will lead to predominantly collisional outcomes.  
Furthermore, we 
would like to minimize the computing time, which leads us to consider
closely spaced systems, while avoiding MMRs.  

Since our simulations in this section do not include the effects of
gas, they are not intended to model the early phases of planet
formation.  Instead, at $t=0$, we begin integrating fully formed
planetary systems with a disk sufficiently depleted that the planets
are free to interact with each other without significant dissipation
from the disk.  In \S\ref{overview}, we show that the time until
instability within a particular set of initial conditions does not affect the 
statistical properties of final
outcomes. Indeed, we expect that the chaotic dynamics, both {\em
before\/} and {\em after\/} the first close encounter, results in the
distribution of final outcomes being independent of the instability
timescale. This justifies our choice of a very compact initial
configuration with short instability timescale ($\sim 10^4\,$yr), which 
minimizes the computational cost.
See MW02 and Appendix~\ref{timescale} for further discussion of the dependence of the instability
timescale on $K$.

Initial
eccentricities are drawn from a uniform distribution between
$0$---$0.1$, and orbital inclinations are drawn from a uniform
distribution between $0^\circ$---$10^\circ$ (with respect to the
initial orbital plane of the innermost planet).  To
make sure that we could discern any inclination-dependent effects 
we used a slightly broader range of inclinations than seen in
our solar system. However, in \S\ref{inclination}, we show
numerically that the choice of initial inclinations does not
affect the distribution of final inclinations significantly. 
All initial phase angles are assigned random values
between $0^\circ$--$360^\circ$.

\subsection{Planetary Mass Distributions}
\label{mass_dist}

  Our current understanding of
planet formation remains full of uncertainties and no single prescription
can claim to predict a correct planet mass distribution.  For this reason, we
consider three different prescriptions to construct plausible initial mass
distributions for Jupiter-like planets. In all cases we adopt the standard core-accretion
paradigm and we closely
follow the simple planet formation model described in \citet[][hereafter
KI02]{2002ApJ...581..666K}.  Planet masses depend on the distance of
the planet from the central star through the gas
surface density profile of the protoplanetary disk.

\subsubsection{Mass Distribution 1}
\label{NS-massdist1}

In this prescription, we first assign the planetary core masses $M_{\rm core}$ 
assuming a uniform distribution
between $1$--$10\,M_{\oplus}$. We assume that the cores
accrete all gas within $4$ Hill radii (KI02) to reach a
total mass $M$ at a semimajor axis $a$ given by
\begin{equation}
M = 2\pi a\Delta \Sigma_{\rm gas} + M_{core},
\label{planet_forming1}
\end{equation}
where $\Delta = 8 r_H$ is the feeding zone of the planet core and
$r_H$ is the Hill radius of the planet core, given by
\begin{equation}
r_H = \left(\frac{1}{3}\frac{M_{\rm core}}{M_\star}\right)^{1/3} a.
\label{planet_forming2}
\end{equation}
Here $M_\star$ is the mass of the central star and $a$ is the
orbital radius of the core (assumed to be on a circular orbit). 
The gas surface density in the disk is given by
\begin{equation}
\Sigma_{\rm gas} = f_g \Sigma_1 \left(\frac{a}{1
\rm{AU}}\right)^{-\frac{3}{2}} \, {\rm g}\,{\rm cm}^{-2},
\label{gas_density}
\end{equation}
where the coefficient $f_g=240$ is the assumed gas-to-dust ratio 
(taken from KI02), ${\Sigma_1}$ is the surface density at $1\,$AU, and
the exponent comes from the minimum-mass disk model.  We use $\Sigma_1
= 10$ in this case, which is a little higher than the minimum-mass
Solar nebula value of $7$.  The
choice of $\Sigma_1$ is somewhat arbitrary and motivated to produce
roughly Jupiter-mass planets.  The initial masses of the planets
obtained with this procedure are between about $0.4\, M_J$ and $1.2\, M_J$.  

For this mass distribution we performed a set of $1000$ independent 
dynamical runs.

\subsubsection{Mass Distribution 2}
\label{NS-massdist2}

This is a slight refinement on the previous case
in which we adopt an alternative prescription for
accretion that takes explicitly into account the growing mass of the initial
planetary core.  The initial core masses are chosen as in \S \ref{NS-massdist1} but the
final mass of each planet is now determined using the following equations.
Assuming that an infinitesimal mass $dm$ accreted by a planet of mass $m$
decreases the disk density by $d\Sigma$, we can write
\begin{equation}
dm = -2\pi a n_H r_H d\Sigma,
\label{accretion}
\end{equation}
where $n_H$ is the number of Hill radii over which the mass is accreted.  
The final mass $M$ of a planet at a distance $a$
from the star, starting with a core mass $M_{\rm core}$ can be obtained by
integrating Eq.~\ref{accretion} as follows,
\begin{equation}
\int_{M_{\rm core}}^{M} m^{-1/3} dm = -2\pi n_H a^2 \frac{1}{3M_{\star}^{1/3}} \int_{\Sigma_i}^0 d\Sigma, 
\label{integral}
\end{equation} 
where $\Sigma_i$ is
the initial disk surface mass density.  Solving Eq.~\ref{integral} and replacing $\Sigma_i$ from 
Eq.~\ref{gas_density} 
we find
\begin{equation}
M = \left( \frac{4\pi n_H a^2}{M_{\star}^{1/3} 3^{4/3}} f_g \Sigma_1 \left(\frac{a}{1 AU}\right)^{-3/2} + m_c^{2/3}\right)^{3/2}.
\label{mass_accretion}
\end{equation} 
Here we use $n_H = 8$, i.e., we assume that the core accretes all mass within $4$
Hill radii on either side.  We use the same values for $f_g$ and $\Sigma_1$ from 
\S\ref{NS-massdist1}.  

For this mass distribution we have integrated a smaller set of $224$ systems.

\subsubsection{Mass Distribution 3}
\label{NS-massdist3}

We expect that the final distributions
of different orbital properties may vary significantly with different initial
mass distribution.  To further test this mass dependence, we created a third set
 of systems with
a broader planet mass distribution.  Here
we assign planetary masses exactly as in \S \ref{NS-massdist1}, but the initial core 
masses are chosen differently.  We sample $M_{core}$ from a
distribution of masses between $1$--$100\,M_{\oplus}$ uniform in
$M_{core}^{1/5}$, while assuming again that these cores accrete all gas within $8$
Hill radii.  The exponent in the core mass distribution and
the surface density at $1\,$AU, $\Sigma_1 = 15$, 
are chosen somewhat
arbitrarily with the goal to obtain an initial mass distribution that
peaks around a Jupiter mass but with a tail extending up to several Jupiter masses.  
The choices above produce initial masses
spanning about an order of magnitude, in the range $0.4\,M_J$--$4\,M_J$.  
Moreover, the distribution for higher-mass planets resembles the mass 
distribution of observed exoplanets (see \S\ref{mass_dependence}).  

For this mass distribution we have integrated a set of 500 systems.

\subsection{Numerical Integrations}
\label{NI1}

We integrate each system for $10^7$ yr, which is $2
\times 10^6$ times the initial period of the closest planet ($T_{1,i}$),
and typically much longer than the timescale for the
onset of instability.  We use the hybrid integrator of
{\tt MERCURY6.2} \citep{1999MNRAS.304..793C} and integrate the orbits symplectically
while there is no close encounter, with a time-step of $10$ days, but
switching to a Bulirsch-Stoer (BS) integration as soon as two planets have a close
approach (defined to be closer than $3$ Hill radii).  Runs with poor energy conservation
($|\Delta E/E| \geq 0.001$) with the hybrid integrator are repeated
using the BS integrator throughout with the same $|\Delta E/E|$ tolerance.  
This happens in $\sim 30\%$ of all runs,
but our conclusions are not
affected even if we reject these systems.   We find that, in all
systems, at least one planet is eventually ejected.  Note that, for
three-planet systems, following an ejection the remaining two
planets may or may not be dynamically unstable.  Therefore, we do not
stop the integration following an ejection. Instead, we continue all integrations
for two planets until a fixed stopping time of $10^{7}\,$yr. For systems with two
remaining planets we check for Hill stability using the known
semi-analytic criterion \citep{1993Icar..106..247G}.  In our
simulations about $9\%$ of systems were not provably Hill stable at the
integration stopping time.  We discard those from our analysis. 
When a single planet remains (following a second ejection or when
a collision took place), the integration is of course stopped immediately.  

For systems with two remaining dynamically stable planets the orbits 
can still evolve on a secular timescale (typically $\sim 10^5$--$10^6$ yr for 
our simulated systems) much larger than the dynamical instability timescale 
(\citealt{2006ApJ...649..992A}; see also \citealt{2000ssd..book.....M}, Ch~7).  
 We study these systems 
with two remaining stable planets by integrating the secular perturbation equations 
for a further $10^9$ yr 
with the analytical formalism developed in \citet{2000ApJ...535..385F}.  
Note that the more standard formulation for Solar system dynamics 
\citep{2000ssd..book.....M} is not appropriate for these planetary systems 
because a significant fraction present very high eccentricities 
and inclinations.  We find that for most of our simulated systems our chosen 
integration stopping time effectively sampled the full parameter space (see detailed 
discussion in \S\ref{secular}).  

\begin{figure}
\begin{center}
\plotone{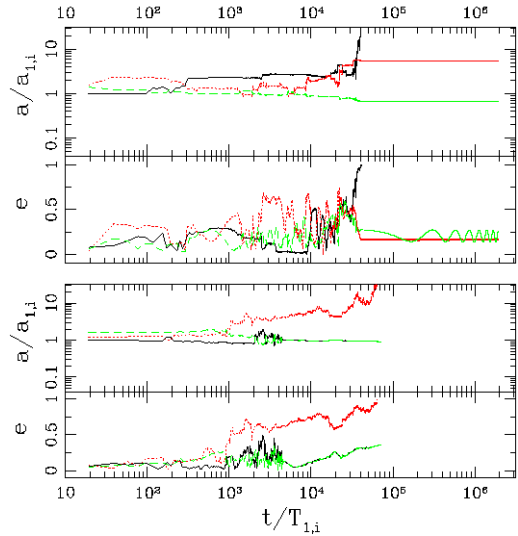}
\caption{Time evolution of semi-major axes and eccentricities 
for two randomly chosen typical simulations.  
The solid (black), dotted (red), and dashed (green) lines show the orbital 
elements for the initially closest ($a_1$, $e_1$), middle ($a_2$, $e_2$), 
and furthest ($a_3$, $e_3$) planets.  The top pair of panels show a 
realization where the first planet is ejected
at $\sim 4.1\times10^4\,T_{1,i}$, and the integration concludes
with two planets in provably stable orbits.  The semi-major axes for
both $P_2$ and $P_3$ remain constant and the eccentricities oscillate
stably on a secular timescale.  The bottom pair shows another realization 
where $P_3$ collides with $P_1$ at 
$\sim 4.2\times10^3\,T_{1,i}$;  $e_2$ keeps 
increasing until, a little before $10^5\,T_{1,i}$ $P_2$ gets ejected, leaving a single 
planet in the system.  Since a single orbit is always stable we 
stop the integration following this ejection.  
Numbers in the subscript represent the
positional sequence of the planets starting from the star and letters
``i" and ``f" mean initial and final values, respectively, in all plots.  
}
\label{time_ej}   
\end{center}
\end{figure}

We treat collisions between planets in the following simple way (``sticky-sphere''
approximation). A collision is assumed to happen when the distance
between two planets becomes less than the sum of their physical radii.
We assume Jupiter's density ($1.33\,{\rm g}\,{\rm cm}^{-3}$) for all 
planets when determining the radius from the mass.  
After a collision the two planets are replaced by a single one
conserving mass and linear momentum.  
Because we account for collisions, our results are not strictly scale
free.  However, we find that collisions are relatively rare for our
choice of initial conditions, so we still present all results with
lengths scaled to $a_{1,i}$ and times scaled to $T_{1,i}$.

Since Mass Distribution 1 corresponds to our largest set of runs, we first show 
our results from this set in detail in the following subsections
(\S \ref{overview} -- \S \ref{MMR}).  Results for the other two sets
 are summarized in \S\ref{mass_dependence}.

\subsection{Overview of Results}
\label{overview}

\begin{figure}
\begin{center}
\plotone{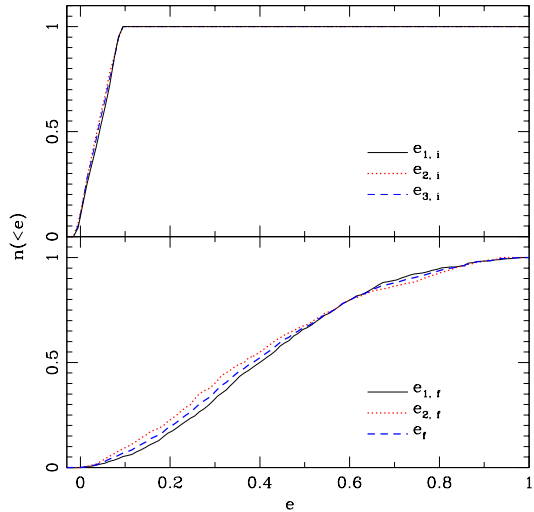}
\caption{Cumulative distributions showing initial and final
eccentricities of the planets.  Top and bottom panels show the initial
and final cumulative eccentricity distributions, respectively.  In the
top panel solid (black), dotted (red) and dashed (blue) lines
represent the closest, middle, and furthest planets, respectively.
They are on top of each other because the initial eccentricity
distribution is the same for all of the planets.  In the bottom panel
solid (black) and dotted (red) lines represent the final inner and 
outer planets, respectively.  The dashed (blue) line shows all remaining 
planets in final stable orbits.  
 }
\label{e}   
\end{center}
\end{figure}

In Fig.~\ref{time_ej} we show a couple of randomly selected, representative examples of the  
dynamical evolution of these systems, showing both chaotic phases and  
stable final configurations.  
Note the order-of-magnitude difference in timescale to first orbit crossing,
illustrating the broad range 
of instability timescales (see also the discussion in Appendix~\ref{timescale}).
We find that strong scattering between planets increases the
eccentricities very efficiently (Fig.~\ref{e}).  The median of the
eccentricity distribution for the final inner planets is $0.4$.  
The median eccentricity for the final outer planets is $0.37$, that
for all simulated planets in their final stable orbits, is $\sim 0.38$.  

\begin{figure}
\begin{center}
\plotone{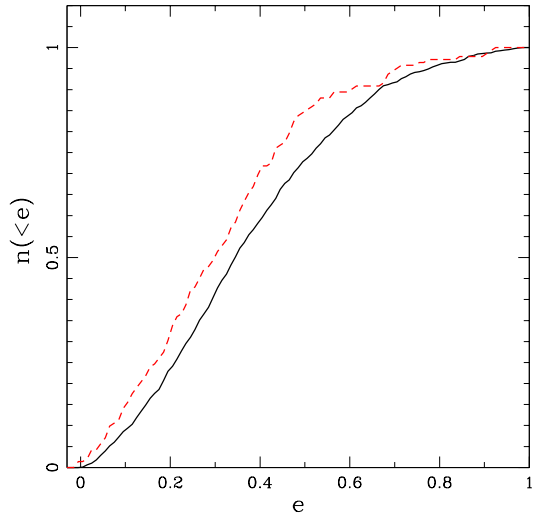}
\caption{Comparison between the simulated and observed 
exoplanet populations.  The solid (black) line shows the cumulative 
distribution of the eccentricities of the remaining planets in their final 
stable orbits.  The dashed (red) line is that for the observed population.  
For this comparison we employ a lower mass cut-off of $0.4\,M_J$ 
on the observed population addressing the fact that we do not have 
lower mass planets in our simulations.  We also consider only the simulated 
planets that are finally within $10\,AU$ from the star to address the fact that 
in the observed population we do not have planets further out.  We also 
employ a lower semi-major axis cut-off of $0.1\,a_{1,i}$ on the observed 
population.  }
\label{e_compare}   
\end{center}
\end{figure}

We compare our results with the observed eccentricity distribution of
detected exoplanets in Fig.~\ref{e_compare}.  For a more meaningful 
comparison we restrict our attention to observed planets
with masses greater than $0.4\,M_J$, similar to the lower mass cut-off
in our simulated systems.  We also place an upper limit on the semimajor axis
at $10\,\rm{AU}$ for the simulated final planet population to
address the observational selection effects against discovering
planets with large orbital periods.  Similarly, since planets close to the
central star can be affected by additional physics  beyond
the scope of this study 
\citep[e.g., tides, general relativistic effects; see][]{2006ApJ...649.1004A} 
we also omit observed close-in planets with
semimajor axes below $0.1\,a_{1,i}$.  

\begin{figure}
\begin{center}
\plotone{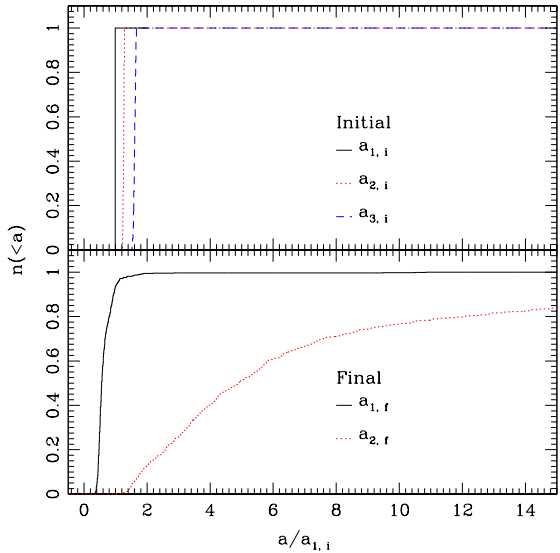}
\caption{Cumulative frequency plots of semi-major axes of the initial (top panel) and 
final (bottom panel) planets.  Vertical solid (black), dotted (red) and dashed (blue) 
lines show the initial values in the top panel.  These are vertical lines because the
initial semi-major axes of the closest, middle and outer planets do not have a spread.  
Solid (black) and dotted (red) curves in the bottom panel
show the final inner and outer planets' semi-major axes, respectively.    
 }
\label{a}   
\end{center}
\end{figure}

As seen in Fig.~\ref{e} and Fig.~\ref{e_compare}, our simulations
slightly overestimate the eccentricities of the planetary orbits.
However, the slopes of the cumulative eccentricity distributions at
higher eccentricity values are similar.  In a realistic planetary
system, there might be damping effects from lingering gas, dust or
planetesimals in a protoplanetary disk.  While our simplified
models already come close to matching the eccentricity
distribution of observed planets, including damping may further improve this
agreement.  
To be more quantitative, we perform 
a Kolmogorov-Smirnov (KS)
test and find that we cannot rule out the null hypothesis (that the
two populations are drawn from the same distribution) at the $85\%$
level (Table~\ref{KS}).  In \S \ref{newmass}, we will show
that a broader initial distribution of planet masses results
in an even better match to the observed eccentricity distribution.  

\begin{figure}
\begin{center}
\plotone{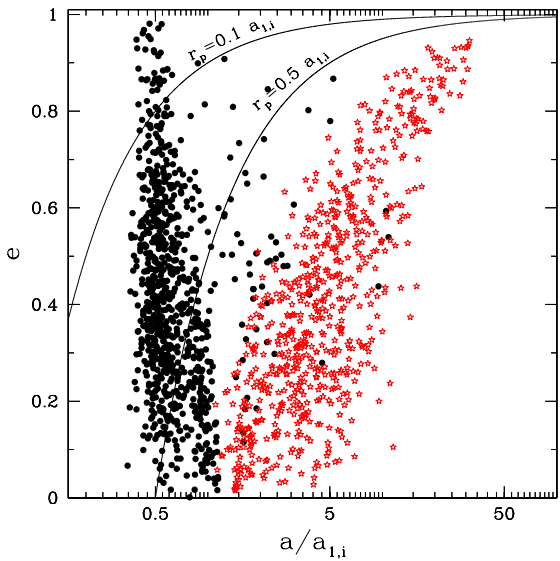}
\caption{Final semi-major axis versus eccentricity plot.  All lengths are
scaled by the initial closest planet semi-major axis (here $a_{1,i} =
3.0 \rm{AU}$).  Black solid circles and red open stars represent the
final inner and outer planets, respectively.  Solid lines show
different constant periapse lines with values 0.1 and 0.5.  Note the
high eccentricities and the close approaches towards the central star.
The empty wedge shaped region in the a-e plane at high eccentricities is
due to the requirement for orbital stability.  
}
\label{a_vs_e}   
\end{center}
\end{figure}

\begin{deluxetable}{ccc}
\tablewidth{0pt}
\tablecaption{Comparison of Eccentricity Distributions \label{KS}}
\tablehead{
\colhead{ } & \colhead{$D$} & \colhead{$P$}}
\startdata
Mass distribution 1 & 0.113 & 0.15 \\
Mass distribution 2 & 0.171 & 0.01\\
Mass distribution 3 & 0.087 & 0.32 \\
\enddata
\tablenotetext{a}{For each mass distribution, we compare the final
eccentricity distribution of the simulated population with the
observed exoplanet population (Figs.~\ref{e_compare},
\ref{e_compare1}).  Using the {\tt{kstwo}} function in {\tt{Numerical
Recipes}}, we calculate the two-sample Kolmogorv-Smirnov statistic,
$D$, and the corresponding probability, $P$.  In each case, the
high value of $P$ indicates that we can not reject the null hypothesis
that both samples were drawn from the same population.
}
\end{deluxetable}

The top and bottom panels in Fig.~\ref{a} show the cumulative
distributions of the initial vs final semi-major axes for the planets.  
The planet that is closest to the star initially may not remain
closest at the end of the dynamical evolution.  In
fact, all three planets, independent of their initial positions, have
roughly equal probability of becoming the innermost planet in the final stable
configuration when the planet masses are not very different.  In $20\%$
of the final stable systems, we find a single planet around the
central star, two planets having been lost from the system either through
some combination of collisions and dynamical ejection.  The other systems 
have two giant
planets remaining in stable orbits.  We find that the planets in the
outer orbits show a tendency for higher eccentricities correlating with larger
 semi-major axes (Fig.\,\ref{a_vs_e}).  We now know that many of the current
observed exoplanets may have other planets in distant orbits
\citep{2007ApJ...657..533W}.  From our results we expect that planets
scattered into very distant bound orbits will have higher eccentricities. Long-term
radial velocity monitoring should be able to test this prediction.  

\begin{figure}
\begin{center}
\plotone{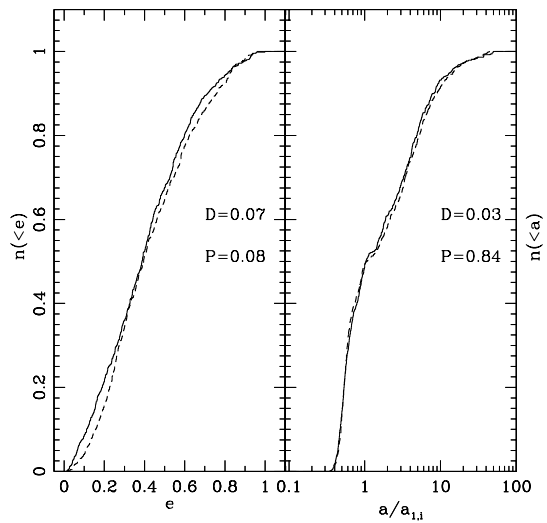}
\caption{Left panel: cumulative frequency plots for the final eccentricities of the two 
subgroups, {\em Group 1\/} (solid line) and {\em Group 2\/} (dashed line).  
Right panel: cumulative frequency plots for the final semi-major axes of the two 
subgroups, {\em Group 1\/} (solid line) and {\em Group 2\/} (dashed line).  KS statistics 
results, $D$ and $P$ are also quoted for each of the above.  $D$ and $P$ are 
as defined in Table~\ref{KS}.  
 }
\label{time_segment_a_e}   
\end{center}
\end{figure}

\begin{figure}
\begin{center}
\plotone{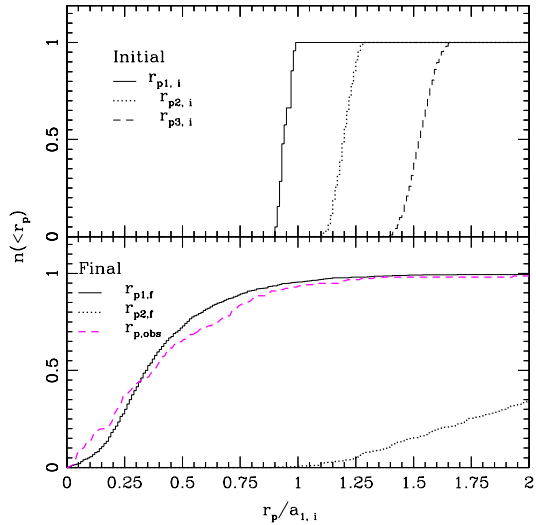}
\caption{Cumulative histogram of the pericenter distance of the
initial (top panel) and final (bottom panel) planets bound to the
star.  In the top panel the solid, dotted and dashed lines show the
pericenter distributions of the initial closest, middle and the
furthest planets, respectively.  In the bottom panel the solid and
dotted lines show the same for the final stable inner and the outer planets
with their semi-major axes less than $10\,\rm{AU}$.  
The dashed magenta line shows the
pericenter distribution of the observed exoplanet population for
comparison purposes.  }
\label{rp_hist}   
\end{center}
\end{figure}

Next, we investigate to what extent the final orbital properties depend
on the instability timescale (equivalently, on how closely packed the initial configuration was).  
For each system we integrated
in \S\ref{NS-massdist1}, we noted the first time when the semi-major axis of any one
of the planets in the system changed by at least $10\%$. We use this as a measure of the
dynamical instability growth timescale.  Then, we
divide the set into two subgroups, based on whether this growth time was
below (Group~1) or above (Group~2) its {\em median\/} value (so 50\% of the
integrated systems are in each group).
Fig.~\ref{time_segment_a_e} compares the final eccentricity and 
semi-major axis distributions between the two groups.  We find that
the distributions are indistinguishable, demonstrating that the final (observable) orbital
properties are not sensitive to when exactly a particular system became
dynamically unstable, as long as the dynamics was sufficiently active (ensuring
that close encouters occur) and avoiding initial conditions so closely packed that
physical collisions would become dominant.
This result is hardly surprising since we expect the chaotic evolution
to efficiently erase any memory of the initial orbital parameters. 
Our results can therefore be taken as representative of the dynamical
outcome for analogous systems with an even larger initial spacing between planets (but avoiding 
mean motion resonances; see Appendix~\ref{timescale}).
In practice, performing a large number of numerical integrations for these more widely spaced initial
configurations would be prohibitively expensive (see Appendix~\ref{timescale}).   

\subsection{Hot Jupiters from Planet--Planet Scattering}
\label{high_e_hot_jup}

  We find that a significant fraction of systems emerge with planets in orbits
having very small periastron distances.  Fig.~\ref{a_vs_e} 
shows the final positions of the planets that are still
bound to the central star in the $a-e$ plane.  The solid lines represent
different constant pericenter distances.  Note that the planets show weak correlations
between the eccentricity and the semi-major axis.  For the inner
planets, a lower semi-major axis tends to imply higher
eccentricity, while the outer planets show an opposite trend.
The final inner and outer planets form two clearly separated clusters
of points in the $a-e$ plane due to stability considerations.

Fig.~\ref{rp_hist} shows the cumulative distribution of the
periastron distances of the final bound planets around the star.  For
the sake of comparison, we also show the pericenter distribution of
all observed exoplanets in Fig.~\ref{rp_hist}.  We see that $10\%$ of the
systems harbor planets with periapse distances $\leq 0.05
a_{1,i}$, whereas, a few ($\sim 2\%$) harbor planets with periapse distances 
$\leq 0.01a_{1,i}$.  Since we do not include tidal effects, we cannot compare 
this quantitatively with the observed population.  However, this is consistent with 
the $\sim 5\%$ of observed planets with semi-major axes within $0.03\,\rm{AU}$.    
If the initial semi-major axes are sufficiently small
tidal forces could then become important and a planet on a
highly eccentric orbit could be circularized to produce a hot Jupiter
\citep{2006ApJ...638L..45F,2005Icar..175..248F,1996Sci...274..954R,
1996Natur.384..619W,2002Icar..156..570M}.  However, recall that
systems with much smaller values of $a_{1,i}$ would also lead to 
more physical collisions than in our simulations.  Moreover, a 
full numerical study of this scenario should include tidal dissipation
as part of the dynamical integrations, and possibly also include 
additional physics such as GR effects, etc.  
\citep{2008arXiv0801.1368N}.

\subsection{Planets on High-inclination Orbits}
\label{inclination}

\begin{figure}
\begin{center}
\plotone{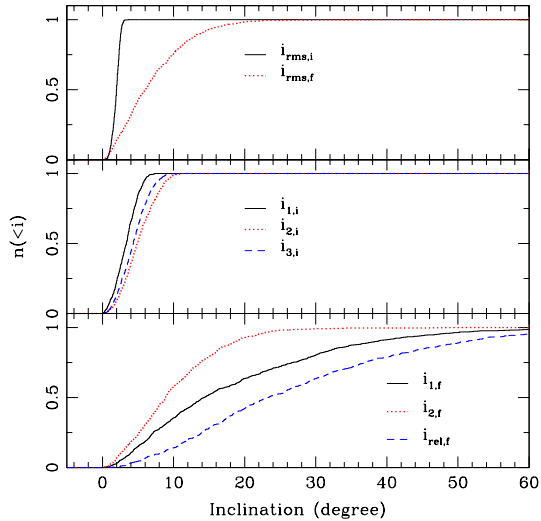}
\caption{Cumulative distribution showing initial and final orbital
inclinations of the planets with respect to the initial invariable
plane.  In the top panel the solid (black) and the dotted (red) lines
represent the initial and final RMS inclination distributions of the
planet orbits with respect to the initial invariable plane.  In the
middle panel solid (black), dotted (red) and dashed (blue) lines
represent the closest, middle and furthest planets, respectively.  The
bottom panel shows the final orbital inclination distributions of the
remaining planets in the system.  The solid (black), and the dotted (red) 
lines represent the inner and outer planets, respectively.  The dashed 
(blue) line represents the relative angles between the two remaining 
planetary orbits.  Note that the final closer planets,
which are the planets more easily observable in a planetary system,
statistically have higher inclinations.  Note, that the relative 
inclinations between the planetary orbits are also quite high.  }
\label{i}   
\end{center}
\end{figure}

Since the star and planets get their angular momenta from the same
source, planetary orbits are generally expected to form in a coplanar disk
perpendicular to the stellar spin axis.  In Fig.~\ref{i}, we compare
the distributions of the final inclination angles.  Here each
angle reported is the absolute value of the orbital inclination
measured with respect to the initial invariable plane,
defined as the plane perpendicular to the
initial total angular momentum vector of the planetary orbits.  Note
that the direction of the initial total angular momentum can differ
from the direction of the total angular momentum of the bound planets
at the end of a simulation, since planets are frequently ejected from
the system, carrying away angular momentum. 

\begin{figure}
\begin{center}
\plotone{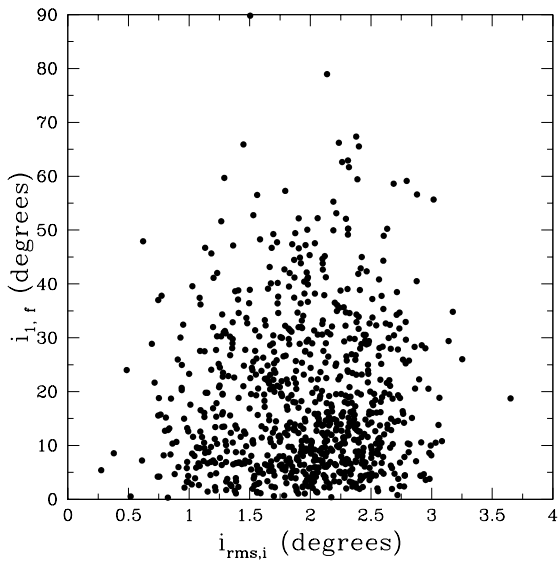}
\caption{Initial RMS inclination vs final
inclination of the inner-most planet.  Note that the final closer
planet orbital inclination is largely insensitive to the initial RMS
inclination.  }
\label{ii}   
\end{center}
\end{figure}

Strong scattering between planets often increases inclinations of
the orbits, leading to a higher final RMS value of planet inclinations compared
to the initial configuration (Fig.~\ref{i}, top panel).  In
general, the inclinations tend to increase for all planets.  The
middle and bottom panels in Fig.~\ref{i} show the initial and 
final inclinations of the orbits of individual planets,
respectively.  The inclination of the final inner planet is typically
larger than that of the final outer planet (Fig.~\ref{i}, bottom
panel).

Our results show that strong planet--planet scattering can dramatically
affect the coplanarity of some planetary systems (Fig.~\ref{i},
bottom panel).  Since the timescale for tidal damping of inclinations
is usually much greater than the age of the stars \citep{2005ApJ...631.1215W},
significantly increased inclinations could be found in some planetary
systems that have gone through strong gravitational scattering phases
in their lifetimes.  Measuring a poor degree of alignment among the
planetary orbits in multiple-planet systems, or between the angular
momentum of one planet and the spin axis of its host star, could be
used to identify systems that have undergone a particularly tumultuous
dynamical history.  

\begin{figure}
\begin{center}
\plotone{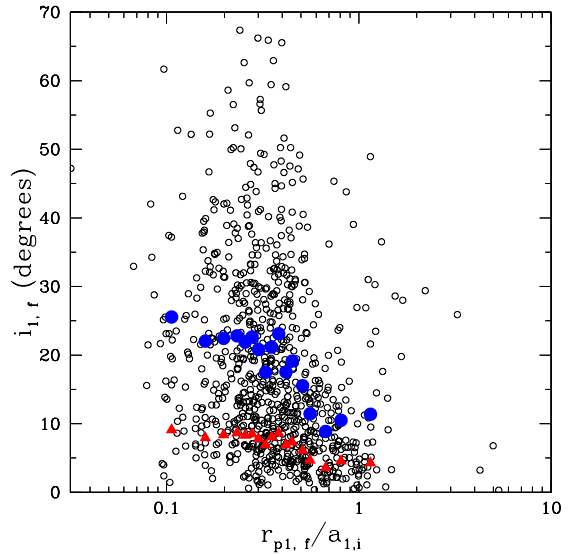}
\caption{Pericenter distance vs inclination of the final inner
planets.  The open dots show the final positions of the final inner planet
in the pericenter-inclination plane.  The filled disks (blue) and triangles (red) 
represent the mean orbital inclination of the inner planet and the final RMS 
inclinations, respectively.  The means are obtained for bins of equal population 
($n_{bin}=50$).  We observe a weak
anti-correlation between the pericenter and the inclination.  }
\label{irp}   
\end{center}
\end{figure}

If a system were initially assigned to a strictly coplanar
configuration, then angular momentum conservation dictates that it
would remain coplanar always.  However, away from this trivial
limit, we expect little correlation between the initial and final
inclinations, given the chaotic nature of the dynamics.  We
test this hypothesis here by investigating the correlation between the
initial and final inclinations of all planets in our simulations.  We find that the
final inclination of the inner planet indeed does not depend on the initial
RMS inclination (Fig.~\ref{ii}).  We can quantify the amount of
correlation between the initial RMS inclination and the final orbital
inclination of the final inner planet using the bivariate correlation
coefficient.  The bivariate correlation coefficient ($r_{xy}$) for two
variables x and y, is given by the following equation,
\begin{equation}
r_{xy}=\frac{\rm{Cov}(x,y)}{sd(x)sd(y)},
\label{rxy}
\end{equation}
where $\rm{Cov}(x,y)$ is the covariance of $x$ and $y$, 
and $sd(x)$ or $sd(y)$ is the standard 
deviation of x or y.  
We find that the correlation coefficient between the initial RMS and
the final inner planet orbital inclinations is $r_{iRMS,iclose} = 0.05$.  
The low
value of $r$ confirms that the high final inclinations are
not merely a reflection of the initial conditions.  As long as
the planetary system is not strictly coplanar initially, strong
planet--planet scattering can increase the orbital inclinations of some
systems significantly.

The final inclination of the inner planet, which is the most
easily observable, shows a weak anti-correlation with the pericenter
distance of its orbit (Fig.~\ref{irp}): lower pericenter orbits
tend to have higher inclinations.  The correlation coefficient in this
case is $r_{rp,iclose}=-0.13$ (Eq. \ref{rxy}).  

For our solar system, the angle between the spin axis of the Sun
and the invariable plane is $\simeq6^{\circ}$.  The angle between the
stellar rotation axis and the orbital angular momentum of a transiting
planet ($\lambda$) can be constrained via the Rossiter-McLaughlin
effect.  Observations have measured $\lambda \sin i$ for five systems
(Winn 2006b): $-4.4^{\circ}\pm1.4^{\circ}$ for HD 209458b
\citep{2005ApJ...631.1215W}, $-1.4^{\circ}\pm1.1^{\circ}$ for HD
189733b \citep{2006ApJ...653L..69W}, $11^{\circ}\pm15^{\circ}$ for HD
149026b \citep{2007ApJ...667..549W}, $30^{\circ}\pm21^{\circ}$ for TrES-1b
\citep{2007astro.ph..2707N}, and, most recently, $62^{\circ} \pm 25^{\circ}$ for 
HD 17156b \citep{2007arXiv0712.2569N}.  Our study implies that planetary systems
with a tumultuous dynamical history will sometimes show a large
$\lambda$.  Therefore, we look forward to precise measurements of
$\lambda$ for many planetary systems to determine the fraction of
planets among the exoplanet population with a significant inclination.  
In particular HD 17156b is very interesting in this regard, since the 
potentially high $\lambda$ together with the high eccentricity ($e=0.67$) strongly 
indicates a dynamical scattering origin for this planet.
Measurements of $\lambda$ would be particularly interesting for
the massive short-period planets ($m>M_J$), the very-short period
giant planets ($P<2.5\,\rm d$), or the eccentric short period planets,
since these planets might have a different formation history than the
more common short-period planets with $m\simeq 0.5 M_J$ in nearly
circular orbits.  

\subsection{Mean Motion Resonances}
\label{MMR}

The radial-velocity planet population currently includes 20
multi-planet systems and at least 5 of those systems are in MMR (4
appear to be in a 2:1 MMR).  MMRs can have strong effects on the
dynamical evolution and stability of planetary systems.  The 2:1
MMR is particularly interesting given the proximity of the two orbits
and the increased possibility for close encounters that could result
in strong gravitational scattering between the two planets
\citep{2006A&A...451L..31S,2007A&A...472..981S}.

It is widely believed that MMRs between two or more planets in a
planetary system arise naturally from migration. Convergent migration in a dissipative disk can lead to 
resonant capture into a stable MMR, particularly the 2:1 MMR \citep{2002ApJ...567..596L}.  
Simulations including an empirical dissipative force show that 
planetary orbits predominantly get trapped in 2:1 MMR 
\citep{2005Icar..178..517M, 2008arXiv0801.1368N}.

While we regard differential migration as a natural way to trap
planets into MMRs, we did explore the possibility of trapping two
planets into 2:1 MMR using only the mutual gravitational perturbations
and without any damping.  We certainly expect this to be more difficult
than with damping.  Finding even a
few systems trapped in MMR without any dissipation would be both surprising
and interesting.  In a three-planet system it is possible that one planet
acts as a source or sink of energy to let the other two planets
dynamically evolve into or out of a MMR.  If pure dynamical trapping
into MMRs were efficient, then this would open up interesting
possibilities.  For one, it does not require a common disk origin, as
is a requirement for the migratory origin of MMRs.  Additionally, this
mechanism could operate in a planetary system at a much later time
after the protoplanetary disk has been dissipated.

\begin{figure}
\begin{center}
\plotone{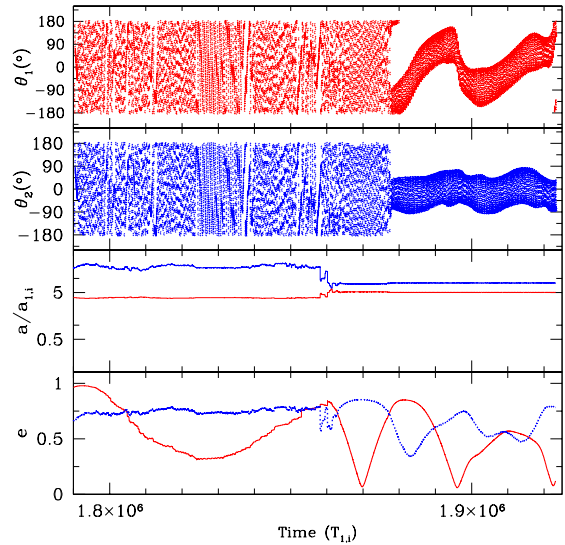}
\caption{Time evolution plots for the two resonance angles $\theta_1$
and $\theta_2$, the semi-major axes and the eccentricities of the
planets.  From top to bottom the panels show the time evolutions of
$\theta_1$, $\theta_2$, semi-major axes and eccentricities,
respectively.  The time axis is in units of the initial orbital period
of the initially closest planet ($T_{1,i}$).  For the panels showing semi-major axes and 
eccentricity, the solid (red) and dotted (blue) lines
show the evolutions of the two planets that enter a 2:1 MMR.  Note that
a little before $1.88\times10^6\,{T_{1,i}}$ both $\theta_1$ and
$\theta_2$ start librating.  
}
\label{reso}   
\end{center}
\end{figure}

To look for possible 2:1 MMR candidates, we isolate systems that
have two remaining planets with their final periods close to a 2:1 ratio.
 Then we calculate the two
resonance angles $\theta_1$ and $\theta_2$ over the full time of their
dynamical evolution.  Here the two resonance angles are given by
\begin{equation}
\theta_{1,2} = \phi_1 - 2\phi_2 + \varpi_{1,2},
\label{theta1}
\end{equation}
where $\phi_1$ and $\phi_2$ are the mean longitudes of the inner
and outer planets and $\varpi_1$ and $\varpi_2$ are the longitudes of
periastron for the inner and outer planets, respectively.  When the planets are not
in a MMR, $\theta_{1,2}$ circulate through $2\pi$.  When trapped in a
MMR, the angles librate around two values \citep{2004ApJ...611..517L}.
Finally, we check whether the periodic ratio and libration of the
resonant angles are long lived or just a transient stage in their
dynamical evolution.  

We find one system where two planets are clearly caught into a 2:1 MMR
(Fig.~\ref{reso}).  The top two panels show the time evolution of the
resonant arguments $\theta_1$ and $\theta_2$.  The two resonant angles
go from the circulating phase to the librating phase at around
$1.88\times10^6\,T_{1,i}$.  The two bottom panels show the
evolution of the semi-major axes and the eccentricities of the two
planets in MMR.  Note that the semi-major axes are nearly constant and
the eccentricities oscillate stably.  Since there is no damping in the
system, the somewhat large libration amplitude of the resonant angles
is to be expected.  In principle, the presence of even a little
damping (due to some residual gas or dust in the disk) might reduce
the amplitudes of libration and eccentricity oscillations for systems
such as this one.  A case like the one illustrated in
Fig.~\ref{reso} is clearly not a typical outcome of purely dynamical
evolution.
We found a few other systems ($\sim 1\%$) showing similar librations of
$\theta_{1,2}$ at different times during their dynamical evolution,
but only for a brief phase  never exceeding $\sim10^4\,T_{1,i}$.  However, if our 
simulations had included even
some weak dissipation, the frequency of such resonances might have
increased significantly.  We encourage future investigation of this
possibility.

\subsection{Mass Dependences}
\label{mass_dependence}

\begin{figure}
\begin{center}
\plotone{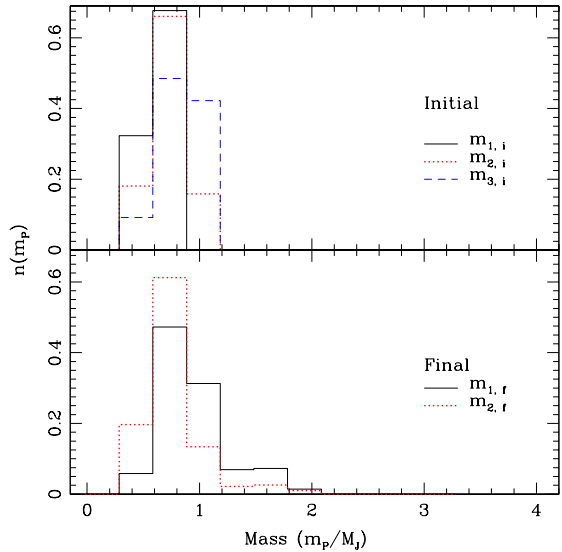}
\caption{Initial and final mass distributions of the closest, middle
and furthest planets.  The top (bottom) panel shows the initial (final) mass
distributions.  Solid (black), dotted
(red), and dashed (blue) lines in the top panel represent the initial
mass histograms of the closest, middle, and furthest planets.  Solid (black) and dotted (red) lines in the bottom
panel represent the mass histograms of the final inner and outer
planets, respectively.  One planet is ejected in
each of our simulations.  Note that the histogram for the inner planet masses
shifts towards higher values in the bottom panel, which indicates that
the higher mass planets preferentially become the inner planet in the
final stable configuration of the planetary systems.  }
\label{m}   
\end{center}
\end{figure}

\begin{figure}
\begin{center}
\plotone{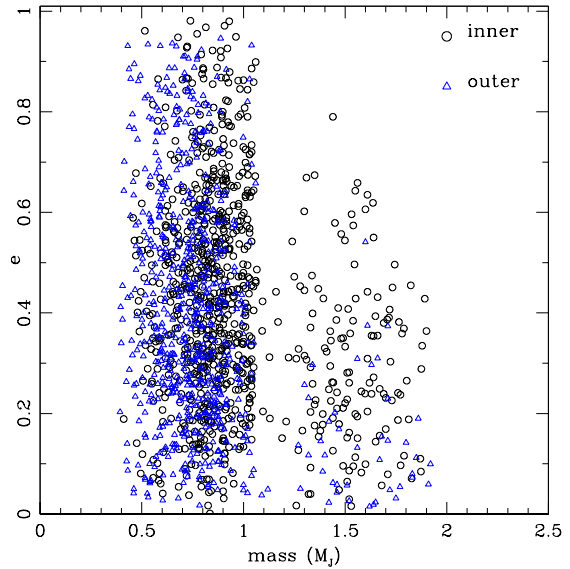}
\caption{Mass vs eccentricity of the final stable planets.  The 
circles (black) and the triangles (blue) represent the final inner and the
outer planets.  Planets with masses $\geqslant 1.6
{M_J}$ are collision products.  The collision planets tend to have lower
eccentricities.  }
\label{m_vs_e}   
\end{center}
\end{figure}

Our simulations show the effects of mass segregation, as heavier
planets preferentially end with smaller semi-major axes.  This trend
can be easily seen by comparing the initial and final mass
distributions of the planets in Fig.~\ref{m}.  The mass
distribution clearly shifts towards higher mass values in the final
inner planet mass histogram, whereas, the outer planet mass
more closely reflects the initial mass distribution (compare the top
and bottom panels of Fig.~\ref{m}).
We do not find a strong effect of mass on eccentricity but we note
that collisions tend to reduce the fraction of highly eccentric
systems (Fig.~\ref{m_vs_e}).  The collision products can be seen in
the cluster around and above $1.5\,M_J$.  We find no other significant
mass dependent effect in the final orbital parameters for our set of
runs using Mass Distribution~1. 

\begin{figure}
\begin{center}
\plotone{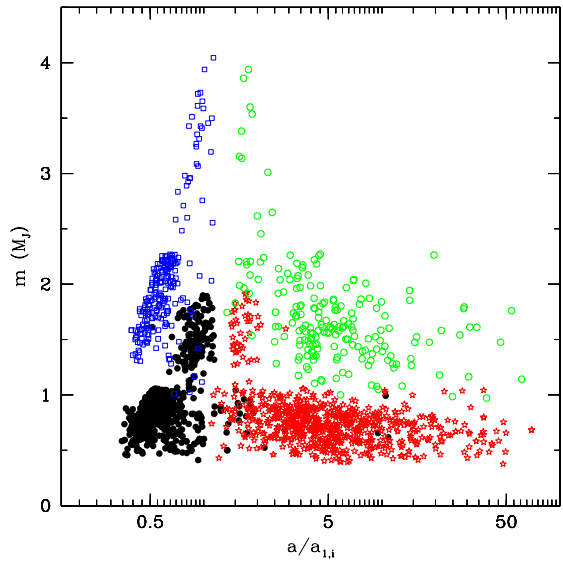}
\caption{Mass vs semi-major axis of the final remaining stable planets.  The 
disks (black) and the open stars (red) represent the final inner and the
outer planets for Mass Distribution~1.  The open squares (blue) and the open circles (green) 
represent the same, respectively, for Mass Distribution~2.  Note, higher mass planets 
remain close to their initial positions.  }
\label{m_vs_a}   
\end{center}
\end{figure}

We now describe briefly the results obtained with the two alternative
initial mass distributions for the three planets.  Fig.~\ref{m_vs_a} shows 
correlation between semi-major axis and mass for both Mass Distribution~1 
(\S\ref{NS-massdist1}) and Mass Distribution~2 (\S\ref{NS-massdist2}).    
Somewhat surprisingly, 
for Mass Distribution~2, we find no significant differences from the results 
obtained with the much simpler prescription of Mass Distribution~1.  
This is possibly because in both Mass Distributions~1 and 2, the mass range and 
distribution are similar (Mass Distribution~2 is only shifted towards slightly higher values).  

\begin{figure}
\begin{center}
\plotone{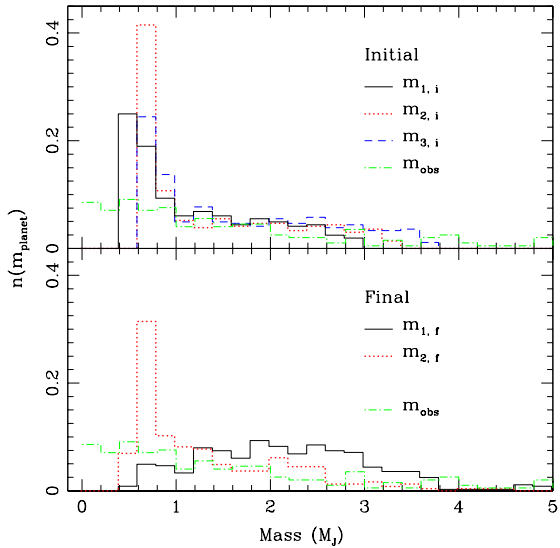}
\caption{Same as Fig.~\ref{m}, but with a different and wider
initial mass distribution than the previous one (Mass Distribution~3).  
The initial mass distribution has a high number
of Jovian mass planets as in the previous mass distribution, however,
in this case the distribution has a tail towards higher masses.  The
higher end in the initial mass spectra in this case mimics the minimum
mass ($m\sin i$) spectrum of the observed exoplanets.  Note that the
mass segregation effect is more prominent here than in Fig.
~\ref{m}.  The dot-dash (green) line shows the $m\sin i$ distribution
of the observed exoplanets in both panels for comparison.  }
\label{m1}   
\end{center}
\end{figure}

\begin{figure}
\begin{center}
\plotone{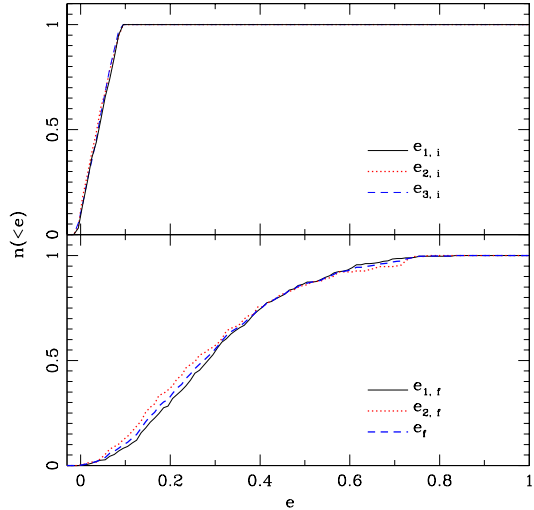}
\caption{Same as Fig.~\ref{e}, but using Mass Distribution~3.  Note that 
overall the eccentricities are lowered using the broader mass distribution.  
}
\label{e1}   
\end{center}
\end{figure}

\begin{figure}
\begin{center}
\plotone{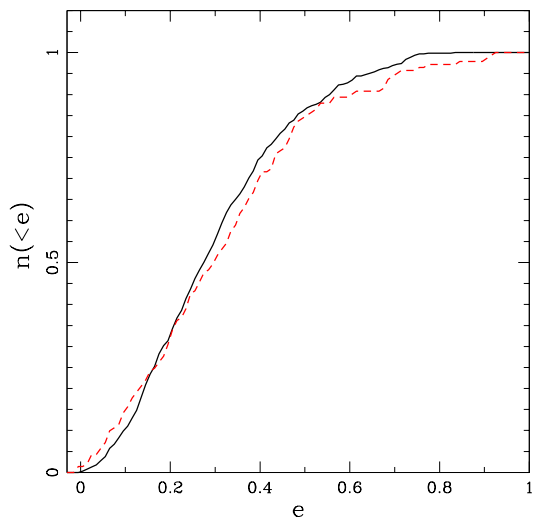}
\caption{Same as Fig.~\ref{e_compare}, but using Mass Distribution~3.  
The simulated eccentricities match much better with the observed in this 
case compared to those using Mass Distribution~1.  
}
\label{e_compare1}   
\end{center}
\end{figure}

\begin{figure}
\begin{center}
\plotone{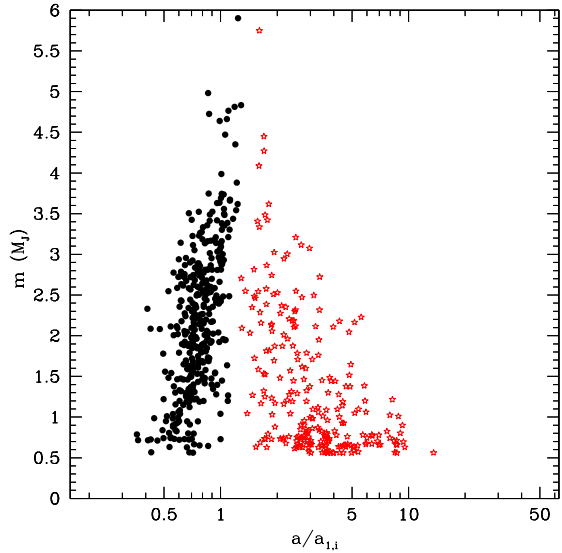}
\caption{Same as Fig.~\ref{m_vs_a}, but using Mass Distribution~3.  Mass 
dependent effects on final semi-major axes of the planets is much more prominent 
in this case.  
}
\label{newmass_m_a}   
\end{center}
\end{figure}

\begin{figure}
\begin{center}
\plotone{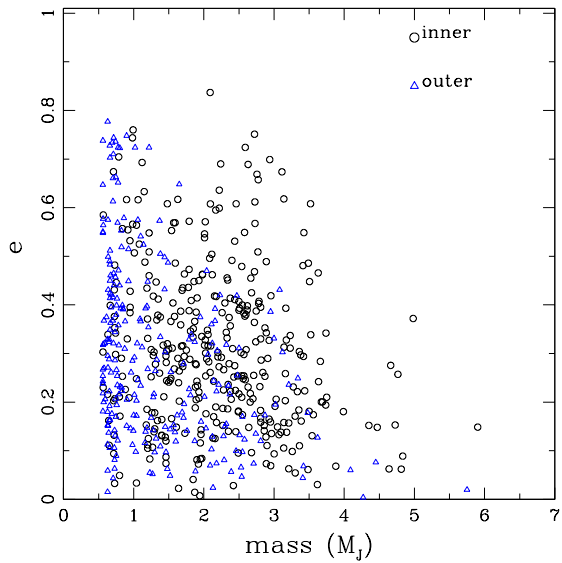}
\caption{Same as Fig.~\ref{m_vs_e}, but using the broader 
distribution of initial planet masses (mass distribution 3).  There seems
to be a weak anti-correlation between the mass and the eccentricities
of the planets.  }
\label{m1_e1}   
\end{center}
\end{figure}

For this reason we also studied a third choice of mass distribution,
Mass Distribution~3 (\S \ref{NS-massdist3}), with a much larger range of planetary 
masses, 
 enabling us to observe mass-dependent effects more clearly.  For example, we 
 now see that the tendency for higher
mass planets preferentially to become the final inner planets
(Fig.~\ref{m1}) is more prominent than in our other simulations.
Similarly, the effect of a mass distribution on the final
eccentricities of the remaining planets is more prominent with this broader
mass distribution.  The higher-mass planets preferentially excite the
eccentricities of the lower-mass counterparts, often to the point of
ejection.  This effectively reduces the overall eccentricities of the
final stable orbits (Fig.~\ref{e1}).  The median value of the final inner orbit eccentricities 
is $0.24$, and that for the outer orbit is $0.23$ in this case.  The final cumulative distribution of
eccentricities matches the observations even more closely
 with Mass Distribution~3 than with Mass Distribution~1 or~2 (Fig.~\ref{e_compare1}; 
 Table~\ref{KS}).  We employ similar selection criteria as described in
\S\ref{high_e_hot_jup}.  
The final semi-major axis distribution is statistically
indistinguishable from the one obtained with
Mass Distribution~1.  We also clearly see that the lower-mass planets
get scattered around preferentially while the heavier counterparts do
not move much and stay mostly near their initial positions
(Fig.\,\ref{newmass_m_a}).  This is in accord with the
observation that close-in planets are often of lower mass than 
planets with moderate semi-major axes \citep{2008arXiv0803.3357C,2005ESASP.560..833N}.  
At present, the correlation
between planet mass and orbital period for radial-velocity planets is
consistent with a population of systems where the less massive planets
have been scattered inwards.  We predict that planet searches
sensitive to longer-period planets will eventually find a population of
sub-Jupiter planets that have been scattered outwards.  Furthermore, our
simulations predict a negative correlation between mass and orbital
period among such long-period planets, if they are launched
into their current orbits via strong gravitational scattering.
We find that mass and eccentricity have a weak anti-correlation 
(Fig.~\ref{m1_e1}).  We do not find any systems with two planets trapped in
2:1 MMR for this case.  
\label{newmass}     

\subsection{Secular evolution}
\label{secular}
\begin{figure}
\begin{center}
\plotone{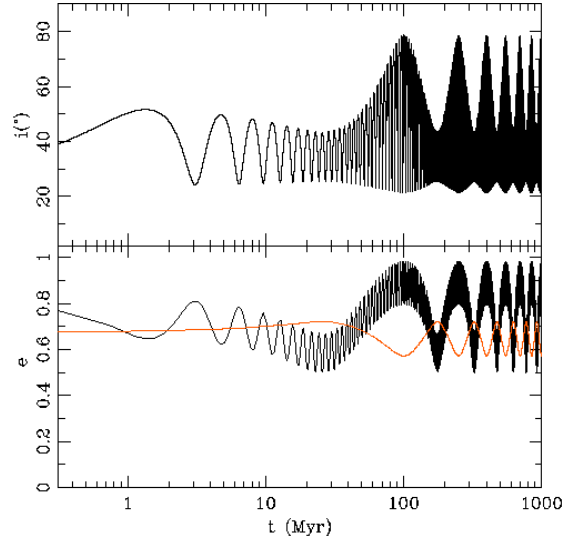}
\caption{Secular time evolution of the relative inclination (top panel) and the eccentricities 
(bottom panel) of a system with two dynamically stable planets.  $t=0$ for this is the end of 
dynamical integration (\S\ref{NI1}).  In the bottom panel the curve with a smaller oscillation 
amplitude (red) shows the evolution of the outer planet eccentricity and the other (black) 
shows that of the inner planet.  }
\label{sec_time_e_i}   
\end{center}
\end{figure}

It is known from numerous previous studies that secular perturbations of one planet on another 
in a multi-planet system can modify the planets' orbital properties on a timescale 
much longer than the relevant dynamical (orbital, or strong dynamical instability) timescales 
(\citealt{2006ApJ...649..992A}; see also \citealt{2000ssd..book.....M}).  
Since secular timescales can be orders of 
magnitude longer than the orbital timescales, 
one might obtain results biased towards the initial part of the 
oscillations if at least a full secular period is not 
sampled properly.  Fig.~\ref{sec_time_e_i} shows a dramatic example 
where the eccentricities of both 
planets and the relative inclinations between the planetary orbits oscillate secularly 
with a very long period ($\sim 100\,\rm{Myr}$) compared to the orbital timescale 
and the observed eccentricities and inclinations 
can be very different from 
what would be expected right after dynamical stabilization of the system.   
Hence, any study of orbital 
properties of planets after dynamical interactions should also worry about the 
secular evolution of the orbital properties that follows the orders of magnitude quicker 
dynamical phase.  Nevertheless, 
we should point out that in our simulated systems this is not typical.  
For most cases the secular time period is 
typically $\sim 10^5$ -- $10^6$ yr. 
For our simulated systems containing two provably stable planets 
at the end of our common integration stopping time ($10^7$ yr) we study the 
evolution of the eccentricities for a further $10^9$ yr to confirm that the orbital 
properties at the end of our integration correctly represent the true final distribution.  

To evaluate the secular evolution of these planets, we use the octupole-order 
formalism presented 
by \citet{2000ApJ...535..385F}.  Note that the more standard formulation 
in terms of the Laplace coefficients \citep{2000ssd..book.....M} is not appropriate 
for these planetary systems because a significant fraction of these systems contain 
orbits with very high eccentricities and inclinations.  

\begin{figure}
\begin{center}
\plotone{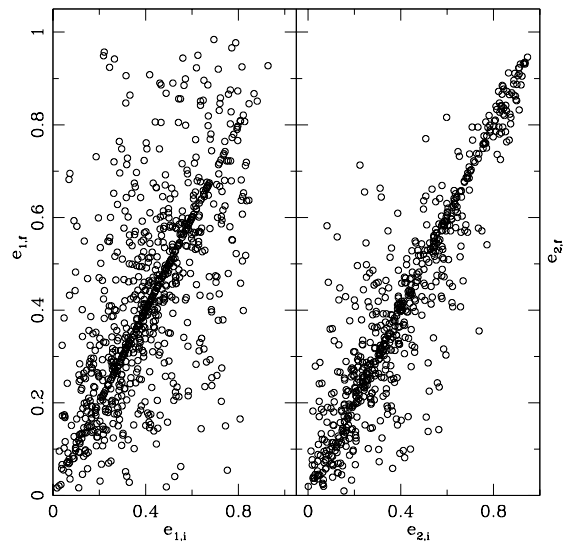}
\caption{Scatter plot of eccentricity after secular evolution for $10^9$ yr vs eccentricity 
after the integration stopping time (\S\ref{NI1}).  Left and right panels show the inner and outer 
planets, respectively.  Note that the eccentricities for the inner orbits change more significantly 
than for the outer ones.  }
\label{sec_e_e}   
\end{center}
\end{figure} 

\begin{figure}
\begin{center}
\plotone{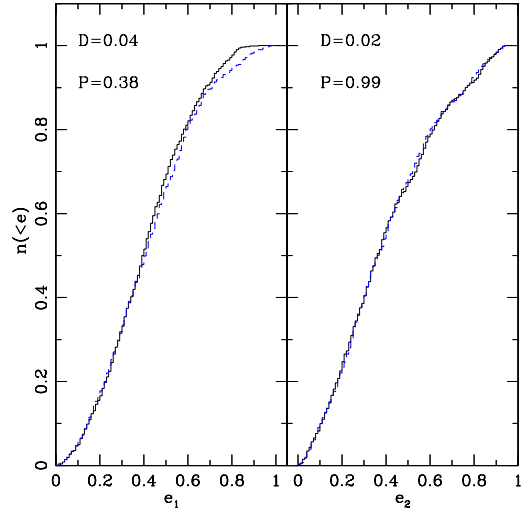}
\caption{Left panel: Cumulative distributions of eccentricities of the inner planets before 
and after secular evolution described in \S\ref{secular}.  Solid (black) and dashed (blue) 
lines show the distributions before and after secular evolution, respectively.  Right panel: 
Same as left panel, for the outer planet.  KS test results for both pairs of distributions are 
also shown in the plot.  }
\label{sec_e_hist}   
\end{center}
\end{figure} 

\begin{figure}
\begin{center}
\plotone{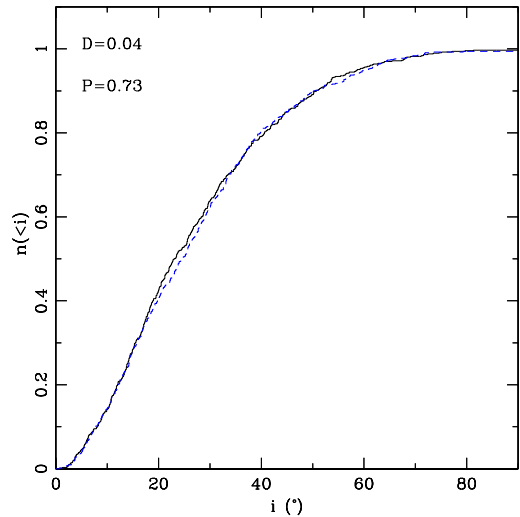}
\caption{Left panel: Cumulative distributions of relative inclinations between the planetary 
orbits before and after secular evolution described in \S\ref{secular}.  Solid (black) and dashed (blue) 
lines show the distributions before and after secular evolution, respectively.  KS test results for 
the two distributions (before and after secular evolution) are also shown in the plot.  }
\label{sec_i_hist}   
\end{center}
\end{figure}

We find that indeed individual eccentricities of these planetary orbits can change significantly.  
Fig.~\ref{sec_e_e} shows a scatter plot of the final eccentricities after secular evolution for $10^9$ 
yr as a function of the eccentricities after our integration stopping time for both planets.  
It is clear that the individual eccentricities can change significantly, especially, for the 
inner planet.   
However, the overall distribution does not change significantly from the distribution obtained 
right after our integration stopping time in \S\ref{NI1}.  Fig.~\ref{sec_e_hist} shows that 
the eccentricity distributions before and after secular evolution for the outer planet, 
in particular, are statistically identical.  For the inner planets we find that, 
after secular evolution, there is a little overabundance of very high eccentricity 
($e\geqslant 0.8$) orbits (Fig.~\ref{sec_e_hist}).  In order to quantify the likeness of 
the two distributions before and after secular evolution, we 
perform KS tests for both the inner and outer planet eccentricity distribution.  We find that we cannot 
rule out the null hypothesis (that the distributions before and after secular evolution are drawn 
from the same distribution) at $62\%$ and $1\%$ significance level for the inner and outer 
planetary orbits, respectively.  The very low values of the significance level along with the large ensemble essentially means that the two distributions are very similar.  We perform the same test 
with the relative inclination of the planetary orbits in the subset of our systems with two 
dynamically stable remaining planets (Fig.~\ref{sec_i_hist}).  For these distributions the 
significance level for KS test with the same null hypothesis is $27\%$.  This confirms that 
our choice of integration stopping time already sampled the full parameter space for 
the secular evolution.  

\section{Effects of a Residual Gas Disk}
\label{gas_planet}

In the previous section we considered the dynamical evolution of
three-planet systems with fully formed planets on initially near-circular 
orbits and no gas disk.  Implicit assumptions are that sufficiently massive
disks damp planetary eccentricities, and that residual gas disks
dissipate quickly enough to allow the later chaotic evolution of planetary
systems. Here, we will verify these assumptions by simulating
three-planet systems within residual gas disks.

\subsection{Photoevaporation}
\label{photoevaporation}

The final stage of disk dissipation remains poorly understood.  
Since viscous evolution alone cannot explain the observed
rapid dispersal of disks 
\citep[$\sim 10^5\,\rm{yr}$; see e.g., ][]{1995ApJ...450..824S}, some other mechanism 
must be responsible for removing a residual disk.  The most likely is photoevaporation
\citep[e.g., ][]{1993Icar..106...92S,1994ApJ...428..654H}.
\cite{2001MNRAS.328..485C} proposed that, once the viscous accretion rate
drops to a level comparable to the wind mass loss rate, photoevaporation takes
over the disk evolution.  When this limit is reached, surface layers
of the disk beyond the gravitational radius ($R_g=GM/c_s^2$),
where the sound speed $c_s$ exceeds the disk's escape speed, starts removing
disk mass faster than it is being replenished by viscous
evolution.  As a result, the disk is divided into inner and outer
parts: the inner disk drains onto the central star on a short viscous
timescale, while the outer disk evaporates on longer timescales
\citep[e.g., ][]{2001MNRAS.328..485C,2006MNRAS.369..229A}.
\citet{2006MNRAS.369..229A} showed that the disk clearing by this
mechanism takes about $10^5\,\rm{yr}$, which is comparable to the observed
dissipation time.  

\begin{figure}
\begin{center}
\plotone{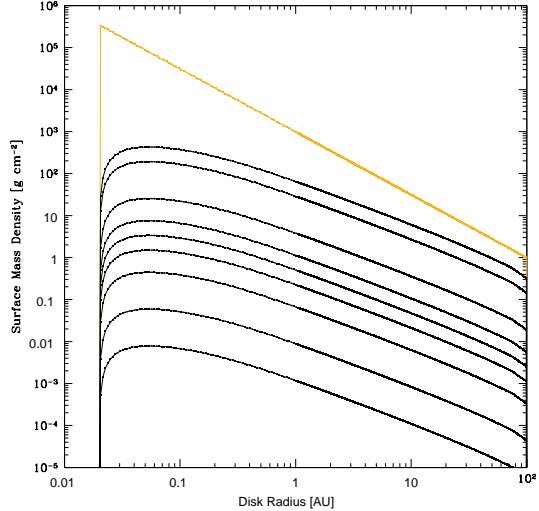}
\caption[surfacemass]{Surface mass densities for runs with DISK1--9 (black line).  
Also shown is the standard surface mass density \(\Sigma=10^3(r/AU)^{-3/2}\) 
(yellow line). The surface mass densities for DISK1--9 are obtained by evolving a 
disk with the standard surface mass density under disk's viscosity parameter 
of \(\alpha=5\times 10^{-3}\) for 3, 5, 10, 13, 15, 17, 20, 25, and 30 Myr.  
\label{surface_gas_dens}}
\end{center}
\end{figure}

The viscous evolution time at semi-major axis $a$ is defined as
\begin{eqnarray}
t_{\rm vis}(a)&=&\frac{M_{\rm disk}\left(\leq a\right)}{\dot{M}_{\rm
disk}\left(a\right)} \\
\dot{M}_{\rm disk}&\simeq& 3\pi\nu\Sigma \ ,
\end{eqnarray}
where \(\nu\) and \(\Sigma\) are the viscosity and surface mass density, respectively.

On the other hand, the photoevaporation time at $a$ is
\begin{equation}
t_{\rm photo}(a)=\frac{M_{\rm disk}\left(\leq a\right)}{\dot{M}_{\rm
wind}\left(a\right)} \ ,
\end{equation}
where the wind mass loss rate for an optically thick disk is
\citep{2001MNRAS.328..485C}
\begin{equation}
\dot{M}_{\rm wind}= 4.4\times 10^{-10} \left(\frac{\Phi}{10^{41} \
{\rm s^{-1}}}\right)^{1/2}\left(\frac{M_*}{M_{\odot}}\right)^{1/2} \
M_{\odot} {\rm yr^{-1}} \ ,
\end{equation}
and for an optically thin disk \citep{2006MNRAS.369..229A}
\begin{eqnarray}
\dot{M}_{\rm wind}= 9.68\times 10^{-10} \mu\left(\frac{\Phi}{10^{41}
\ {\rm s^{-1}}}\right)^{1/2} \left(\frac{h/a}{0.05}\right)^{-1/2} \nonumber \\
\left(\frac{a_{\rm in}}{3 \ {\rm AU}}\right)^{1/2}
\left[1-\left(\frac{a_{\rm in}}{a_{\rm out}}\right)^{0.42} \right] \
M_{\odot} {\rm yr^{-1}} \ .  
\end{eqnarray}
Here \(\Phi\) is the ionizing flux from the central star, \(h\) is
the pressure scale height of the disk, \(a_{\rm in}\) and \(a_{\rm
out}\) are the inner and outer disk radii.

Photoevaporation becomes effective when \(t_{\rm vis}\geq t_{\rm
photo}\).  For typical disks, this corresponds to a disk mass of a few Jupiter
masses. When a disk mass drops below this critical value, planets
are likely to become dynamically unstable if the photoevaporation time
is shorter than the dynamical instability growth time ($t_{\rm
photo}<t_{\rm dyn}$).  In this section we will investigate this
further by simulating 3-planet systems with various disk masses.

\subsection{Numerical Method and Assumptions}
\label{gas_NI}

For this study we use a hybrid $N$-body and 1-D gas dynamics code to
follow the evolution of three-planet systems for several different
disk masses. Our hybrid code in its current form combines an
existing $N$-body integrator with a 1-D implementation of a viscous,
nearly Keplerian gas disk \citep{2005ApJ...626.1033T}.  The $N$-body
code is based on {\tt SyMBA} \citep{1998AJ....116.2067D}. It is fast
for near-Keplerian systems, requiring only $\sim 10$ timesteps per
shortest orbit, while undergoing no secular growth in energy error.
In addition, it makes use of an adaptive timestep to resolve close
encounters between pairs of bodies.

\begin{deluxetable}{cccc}
\tablewidth{0pt} \tablecaption{Disk models \label{disk_model}}
\tablehead{ \colhead{ } & \colhead{Disk Mass} & \colhead{Disk Age ($10^7$\,yr)} 
& \colhead{$t_{damp}$ ($10^6$\,yr)}} \startdata
DISK1 & $3.7\,M_J$ & $0.3$ & $0.2$ for 17/24 \\
DISK2 & $2.2\,M_J$ & $0.5$ & $0.4$ for 14/23 \\
DISK3 & $0.22\,M_J$ & $1.0$ & $4$ for 17/28 \\
DISK4 & $23.5\,M_{\oplus}$ & $1.3$ & $7$ for 5/27 \\
DISK5 & $11.8\,M_{\oplus}$ & $1.5$ & - \\
DISK6 & $4.7\,M_{\oplus}$ & $1.7$ & - \\
DISK7 & $1.4\,M_{\oplus}$ & $2.0$ & - \\
DISK8 & $0.19\,M_{\oplus}$ & $2.5$ & - \\
DISK9 & $0.02\,M_{\oplus}$ & $3.0$ & - \\
\enddata
\tablenotetext{a}{The disk masses of the $9$ different disk models
used in \S\ref{gas_planet}.  The disk age is the time
until a typical disk with $\alpha=5\times 10^3$ will reach the
corresponding total mass. The eccentricity damping time is the time
to reduce the planetary eccentricities from above $0.1$ to below 
$0.1$, and obtained for systems which did not go through mergers. }
\end{deluxetable}

The gas disk is divided into radial bins, each of which represents
an annulus whose properties (surface density, viscosity,
temperature, etc.) are azimuthally and vertically averaged,
following the general approach of \citet{1986ApJ...307..395L}.
Arbitrary viscosities can be specified through a standard
$\alpha$-parametrization \citep{1973A&A....24..337S}. Though this
disk is explicitly 1-D, the vertical and azimuthal structures are
implicitly included in the model.  For the former, a scale height is
assigned to every annulus. The latter is key to the planet--disk
interactions, which result from the raising of azimuthally
asymmetric structure (spiral density waves) in the disk by the
planet.  This effect is added in the form of the torque density
prescription of \citet{1980ApJ...241..425G}, as modified by
\citet{1997ApJ...482L.211W}, which describes the disk--planet
angular momentum exchange taking place as waves are launched. 
Planetary eccentricities are damped on timescales as in
\citet{1993Icar..106..274W} and \citet{1993ApJ...419..166A}.  Since we do not
take account of the saturation of corotation resonances, which could
lead to the eccentricity excitation by Lindblad resonances
\citep{2003ApJ...585.1024G,2008Icar..193..475M}, the eccentricity
damping considered here is an upper limit.

\subsection{Results: Onset of Dynamical Instability}
\label{gas_results}

For initial conditions, we randomly choose $30$ three-planet systems
from the set using Mass Distribution~1 in \S\ref{NS-massdist1}, and
study their orbital evolution within 9 different disk masses.  
The surface mass density profiles of these disks are shown in
Fig.~\ref{surface_gas_dens}. These are obtained by evolving a
minimum mass solar nebula disk model with a viscosity parameter
$\alpha=0.005$ for various times (without planets).  Disk properties
are summarized in Table~\ref{disk_model}, and we will refer to our
models as DISK1-9 from here on.  We assume that each of these
disks is {\it inviscid} for dynamical runs with planets, meaning that 
type II planet migration is not taken into account.  
However, this should not affect our results significantly since
even the most massive disk (DISK1) contains only $3.7 M_{J}$, which is
comparable to the planetary masses used in our
simulations.  Most of our disks are therefore too small to affect planet
migration.

\begin{figure}
\begin{center}
\plotone{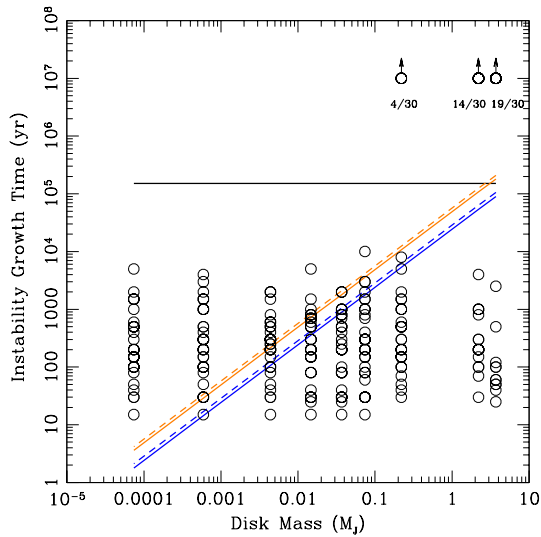}
\caption[3pl_gasdisk]{Instability growth timescale for 30 planetary
systems with DISK1-9 (from right to left).  The range of dynamical
instability time is relatively independent of disk masses.  Data
points with an arrow indicate the number of systems which did not go
through dynamical instability within our simulation time ($\sim
10^7$ yr). Diagonal lines are disk clearing times by
photoevaporation for optically thin disks (dark lines) and optically
thick disks (light lines).  Solid and dashed lines are the disk
clearing time measured at the gravitational radius $R_g$ and $10\,\rm{AU}$
respectively.  The horizontal line shows the viscous evolution timescale of
the disk at $R_g$. Photoevaporation is expected to take over disk
evolution once $t_{\rm viscous}(R_g)\sim t_{\rm photo}(R_g)$.
Planetary systems are likely to go through a chaotic evolution as shown 
in gas free systems (\S\ref{gas_free}) when
$t_{\rm photo}<t_{\rm dyn}$ (i.e. above $t_{\rm photo}$ lines in the
figure). \label{instability_growth}}
\end{center}
\end{figure}

The dynamical instability is commonly characterized by the orbital
crossings of planets.  Fig.~\ref{instability_growth} shows the first
orbital crossing time of each system for each disk mass. 
Diagonal lines are disk clearing timescales by photoevaporation for
optically thick disks (orange/grey lines), and optically thin disks 
(blue/black lines).  Also plotted is the viscous evolution times at the
gravitational radius.  This figure indicates that photoevaporation
takes over disk evolution for disks with a few to several Jupiter
masses, depending on the photoevaporation models.

It appears
that the range of first orbital crossing time $t_{\rm dyn}$ is
relatively independent of disk masses, and around $\sim 10-10^4$ yr.  
Note, however, that the number of systems going through orbital crossings
decreases for larger disk masses. Excluding mergers, $19/24$,
$14/28$, and $4/28$ systems for DISK1, 2, and 3, respectively do not
experience any orbital crossings during the simulation time ($10^7$
yr), while all systems with lighter disk masses go through at least
one crossing within the run time.  This is concordant to expectations that 
planets become dynamically unstable more readily in a less massive 
gas disk and the same planetary system that remained stable in a 
sufficiently massive disk can become unstable once the disk dissipates.  

Apart from $t_{\rm dyn}$, the eccentricity damping timescale ($t_{\rm damp}$) 
is very important to know, since $t_{\rm damp}$ determines 
whether 
the disk can damp the eccentricities 
back to near zero after one (or more) orbit-crossing episode(s), before the disk is 
depleted.  We define $t_{\rm damp}$ as 
the time taken to damp the planetary eccentricities 
from $e>0.1$ to $e<0.1$.  For $30$ different systems 
for $4$ disk masses (DISK1-4) we find the median $t_{\rm damp}$ to be 
$2\times 10^5$, $4\times
10^5$, $4\times 10^6$, and $7\times 10^6$ yr, respectively.   
For less massive disks we do not find significant damping.  
While the evolution of disks is dominated by viscous evolution 
and $t_{\rm damp} < t_{\rm vis}$ (DISK1 and 2),  
planetary orbits are expected
to remain nearly circular since after an instability there is enough time 
to damp the eccentricities before the disk is depleted.  
The nature of evolution can change drastically once 
photoevaporation dominates the disk evolution and starts depleting the disk more rapidly.
We find that most systems reach at least one orbit-crossing episode for
the least massive disks (DISK8 and 9) since $t_{\rm dyn}>t_{\rm photo}$.  
Some planetary systems in more massive disks have $t_{\rm dyn}<t_{\rm photo}$ 
(Fig.~\ref{instability_growth}).  For these more massive disks eccentricities 
excited via planet-planet interaction may be
damped if $t_{\rm damp} < t_{\rm photo}$.  However, since the median 
$t_{\rm damp}$ is longer than $t_{\rm photo}$ for these disks, planetary eccentricities excited
via planet-planet interaction do not have time to be damped before
the gas disk is depleted, once photoevaporation is efficient.

In summary, we expect that planetary systems will remain
stable with nearly circular orbits while the planets are embedded in
a sufficiently massive disk.  Even if
there is an occasional orbital crossing or merger, the
eccentricities and inclinations will rapidly damp in such a disk, so
that the system returns to nearly circular orbits. Then
eccentricities will evolve more freely once photoevaporation takes
over the disk evolution, and the disk clearing time becomes short
compared to the instability growth time ($t_{\rm dyn}>t_{\rm photo}$).
Even when planets become unstable before the disk is completely depleted
($t_{\rm dyn}<t_{\rm photo}$), it is unlikely that their
eccentricities are damped, since the eccentricity damping times of these
disks tend to be longer than the disk dissipation time ($t_{\rm
damp}>t_{\rm photo}$).  Therefore, we expect that most planetary systems
become dynamically unstable when a gas disk dissipates.  This
further justifies our initial conditions in \S\ref{gas_free}.  In a future paper we will further 
investigate the evolution of multiple-planet systems within an
evolving gas disk \citep{Matsumura08}.

\section{Comparison with Previous Studies}
\label{comparison}

The previous work  most similar to ours was the 
pioneering study by MW02 on (gas-free) three-planet systems.  
MW02 also studied the orbital properties of planetary systems following a dynamically active 
phase of their evolution.  However, 
their study was computationally limited and their systems were rather 
idealized in terms of assumed planetary masses and initial  orbits. 
Our results are in good qualitative agreement with those of MW02.  For example, 
they showed for the first time with three-planet systems how scattering can produce 
large eccentricities.  However, 
our more realistic and generalized initial conditions enable us to explore a larger 
parameter space and to study in more detail the most interesting phenomena such 
as the generation of  large, potentially observable inclinations.  We also find that 
the final stable planets can be scattered at even smaller semi-major axes than they 
predicted.  Since these very low semi-major axes planets are in the tail of the distribution, 
it is expected that a simulation of a smaller sample size will miss some of them 
(see Appendix~\ref{size_statistic}).
Moreover, our much larger simulated sets and improved statistics on 
dynamical outcomes allow
us to better compare our theoretical predictions to observations 
(see Appendix~\ref{size_statistic}).  

In addition to the orbital properties of remaining planets, MW02 also presented a stability 
timescale analysis for planetary systems with three giant planets.  In verifying these results, 
we realized the importance of 
this study, especially for our choice of initial spacing, and we therefore decided to perform 
a much more detailed timescale analysis,
with significant  improvements over MW02 made possible by the dramatically increased 
speed of present-day computers.  The results of this analysis are presented in 
Appendix~\ref{timescale}.

\citet{2005Icar..178..517M} studied in detail two--planet systems with an empirical 
dissipation arising from a residual disk.  They found that, even in initially well separated 
two--planet systems, migration can bring the 
planets close enough for dynamical instability.  
In their study they accounted for a disk outside both planets with their empirical formula, 
whereas, we immerse the three planets in a protoplanetary disk with varying 
disk masses (\S\ref{gas_planet}).  Another major difference between their study and 
ours is the number of planets considered.  
The dynamical evolution of two--planet systems can be very different from that of 
systems with three or more planets (see \S\ref{intro}).   
Keeping these differences in mind, we compare key points between the two studies.  
For example, for sufficiently massive disks we find that the 
eccentricity damping timescale is less than the disk dissipation timescale.  However, 
as the disk mass is diminished, the timescale for eccentricity damping and the 
number of unstable systems increases.  They also find that scattering fills up the a--e plane for the 
inner planet orbit.  
Due to 
the setup of their initial conditions and 
the dynamical limitations of two-planet system they do not find planets 
with large orbital periods, normally produced by strong scattering between planets.  They also 
stop integrating after $1\,\rm{Myr}$ or when the system has only one planet left.  One should 
remember that in cases where two planets are remaining, the planetary properties can still 
change either through dynamical scattering (see discussion in \S\ref{NI1}) or even for 
dynamically stable systems, through long term secular perturbations 
(for a detailed discussion see \S\ref{secular}).  

More recently \citet{2007astro.ph..3160J} perform an interesting study as an extension 
of the pioneering work by \citet{1997ApJ...477..781L}.  
They study the dynamical evolution of generic
$N$-planet systems with $N\geq 3$ and a wide range of initial conditions.  
Although their three-planet
systems were dynamically inactive as a result of their choice of initial
separations and integration stopping time, their other runs with higher
$N$ bear very relevant results for our study.  In particular, they find a
similar final eccentricity distribution, suggesting that this distribution
may be universal.  One of the most interesting results in their study is
that the final number of surviving planets following a dynamically active phase is
almost always $2$--$3$, independent of the initial number.  Since these systems are chaotic,
the properties of any planetary system emerging out of a dynamically active phase
will have little memory of the initial number of planets or the exact
initial conditions (also see discussion in \S\ref{overview} and
\S\ref{inclination}).
One can imagine a situation where a system started with $N>3$ and,
followed by many collisions and ejections, reaches a stage with $N=3$.
If dissipation circularizes orbits after each ejection or collision,
then such a system could reach a state similiar to the initial
conditions for our three-planet simulations.  Thus, our results may be 
representative of even more generic multi-planet systems.  

\citet{2008arXiv0801.1368N} study the dynamics of three {\it equal-mass} 
planets including dynamical tides.  Since they can apply 
tides while the three--planet dynamical scattering phase is still active, they 
find increased efficiency to tidally isolate planets that would otherwise still actively 
take part in three--planet scattering.  In this study we did not 
include tides.  However, we find that $~8\%$ of these systems have one planet 
accreted onto the star.  We also find that $\sim 40\%$ of these systems contained 
at least one planet which, during the three--planet violent scattering phase, 
reached a pericenter distance within $0.01\,{\rm AU}$ of the star.  In most of the 
systems these close--in planets end up being ejected, 
while, some of them collide with the star or another planet.  None of these 
systems remain stable at the end of the run.  The addition of 
tides during these scattering phases may stabilize some systems by isolating the 
close--in planet from the other planets dynamically.  Of course the circularization 
process needs to be very efficient and quick so that the planet gets circularized 
and decoupled from the others before it can be ejected.  We should also point out  
that \citet{2008arXiv0801.1368N} study {\it equal-mass} planets.  We find 
that the dynamical evolution of planetary systems with unequal masses is 
very different than for equal mass systems (see discussion in 
\S\ref{mass_dist} \& \S\ref{mass_dependence}).  In particular, the lower-mass 
planets preferentially get scattered inwards or outwards while the heavier 
counterparts remain near their initial positions throughout the whole evolution 
(e.g., see Fig.~\ref{m_vs_a}).  One should also remember that the tidal circularization 
timescale depends on the mass and radius of these planets 
\citep[e.g., ][]{2004MNRAS.353.1161I}.  This can affect  
the efficiency of tidal circularization for high eccentricity planets in their setup.  
At present their study actually produces too many ($30\%$) ``hot" planets 
compared to the current observed population of $\sim 5\%$ within $0.03\,\rm{AU}$ (at 
$0.03\,\rm{AU}$ the tidal circularization timescale is $\sim 10^6\,\rm{yr}$ for Jovian 
planets, see \citealt{2008arXiv0801.1368N}).  
Note that the selection biases of radial velocity surveys can only reduce the 
fraction of hot Jovian planets in the future.  
It will be interesting to see results of similar studies with a more realistic mass distribution.  

\section{Summary and Conclusions}
\label{summary}

We have studied in detail how the orbital properties evolve through
strong gravitational scattering between multiple giant planets in a
planetary system containing three giant planets around a solar mass
star.  
We perform a detailed study for gas free generic planetary systems.
 We focus on the final orbital properties of 
the planets that remain
bound to the central star in stable orbits after chaotic evolution due
to strong mutual interactions, followed by a prolonged secular evolution 
($\sim 10^9\,\rm{yr}$) when two planets remain after the scattering phase.  
We perform the experiments with
realistic planetary systems containing $3$ giant planets (\S\ref{gas_free}).  In
all of these systems at least one planet is eventually ejected
before reaching a stable configuration.  This supports models of
planet formation that predict planetary systems initially form several
closely spaced planets, but instabilities reduce the number of planets
until the stability timescale exceeds the age of the planetary system.
In $20\%$ of the cases, two planets are lost through ejections or
collisions leaving the system with only one giant planet.  Thus, the
planet scattering model predicts the existence of many systems with a
single eccentric giant planet, as well as many free floating planets
(depending on how many planets are formed before the planet scattering
phase of evolution).

We find that strong gravitational scattering between giant planets can
naturally create high-eccentricity orbits.  The exact
distribution of eccentricities for the final remaining planets in
stable orbits depends on the choice and range of the initial mass
distribution.  When the initial mass distribution spans a broad range
of masses, the less massive planets typically start to acquire larger
eccentricities.  However, these planets with highly excited orbits are
often ejected, reducing the overall eccentricities of the remaining
dynamically stable planetary orbits.  Although the first two sets of
our models (with a narrower range of initial planet masses) predict
eccentric planets to be slightly more common than observed
(Fig.~\ref{e_compare}), a wider initial mass
distribution can result in remarkable similarity with the observed
distribution (Fig.~\ref{e_compare1}).  Recently, a similar trend
in eccentricities was found independently by \citet
{2007astro.ph..3160J} for generic dynamically active multi-planet
systems independent of the details of the initial conditions or the
initial number of planets.  

We conclude that planet--planet scattering could easily account for
the observed distribution of eccentricities exceeding $0.2$.  However, our
simulations slightly under-produce systems with eccentricities less
than $0.2$.  This may suggest that some observed systems are affected
by late stage giant collisions.  Alternatively, the presence of a residual gas 
or planetesimal disk could lead to
eccentricity damping.  We find this latter explanation particularly
attractive given the observed correlation between planet mass and
eccentricity \citep{2006ApJ...646..505B}.  While our simulations suggest
that high eccentricities are most common among less massive giant
planets, the known population of extrasolar planets suggest that high
eccentricities are more common among the more massive planets
\citep{2007astro.ph..3163F}.  This apparent discrepancy could be
resolved if a modest disk often remains after the final major
planet-planet scattering event.  Less massive planets would be more
strongly affected by the remaining disk, so their eccentricities could
be damped, while more massive planets would typically be immune to
eccentricity damping.

We find that it is possible to scatter some planets into orbits with
low perihelion distances (Fig.~\ref{rp_hist}).  Approximately $10\%$
of the systems obtain perihelion distances less than $0.05\,a_{1,i}$, 
whereas, a fewer fraction ($\sim 2\%$) can reach within $0.01a_{1,i}$.
If the initial semi-major axes are small enough, then strong
gravitational scattering could result in planet orbits with
sufficiently small perihelion distances, such that tidal effects could
circularize their orbits at small orbital distances.

We find that the inclination distribution of such planets could be
significantly broadened.  If we assume that the angular momentum of
the host star is aligned with that of the initial orbital angular
momentum of the planets, then measurements of $\lambda$ (the angle
between stellar spin axis and planet's orbital angular momentum) should
typically be small in absence of perturbations from other planetary or
stellar companions (\S\ref{inclination}).  We find that strong
gravitational scattering between the giant planets can naturally
increase the inclinations of the final planetary orbits with respect
to the initial total orbital angular momentum plane (Fig.~\ref{i}).
Since the timescale to tidally align the stellar spin and the
planetary angular momentum is much greater than the age of the star
\citep[$\sim 10^{12}\,\rm{yr}$; ][]{1974Icar...23...51G,1980A&A....92..167H,2005ApJ...631.1215W},
inclinations excited by planet-planet scattering after the disk had
dispersed could be maintained for the entire stellar lifetime.
Observations of a hot-Jupiter with a significantly non-zero $\lambda$
would be suggestive of previous planet--planet scattering.  However,
caution would be necessary if the star had a binary stellar companion
\citep{2003ApJ...589..605W,2005ApJ...627.1001T,2007ApJ...669.1298F}.
On the other hand, observations of many hot-Jupiters with orbital
angular momenta closely aligned with their stellar rotation axis would
suggest a formation mechanism other than strong gravitational
scattering followed by tidal circularization. Unfortunately, current
observations measure this angle for only a few systems and some
measurements have uncertainties comparable to the dispersion of
inclinations found in our simulations.  We encourage observers to
improve both the number and precision of Rossiter-McLaughlin
observations.

We find that the relative inclinations between planetary orbits in the
systems with two remaining planets in their final stable orbits also
increase via planet-planet scattering.  Future observations using
astrometry or transit timing could possibly measure relative
inclinations between planetary orbits in multi-planet systems.  
Furthermore, 
we find that in $\sim 20\%$ of the systems having two giant planets in their final 
dynamically stable configurations, the relative inclination between the two planets 
is higher than $40^{\circ}$.  For these systems it is possible for the planets to go through 
Kozai-type oscillations  \citep{2008arXiv0801.1368N}.  
Although effects of a debris disk on planetary dynamics and vice versa
is beyond the scope of this study, the warped disk observed in
$\beta$ Pictoris could be one interesting example where inclined
planetary orbits and the debris disk exchange torques, resulting in a
warped debris disk \citep{1984Sci...226.1421S,2000ApJ...539..435H}.
\citet{1997MNRAS.292..896M} suggests that the observed asymmetry in
the debris disk can be explained by the presence of a planetary
companion in an inclined orbit.  Strong planetary scattering, as
we find, can be a natural way to create planetary orbits with large
semi-major axes and highly inclined orbits.  

Less massive planets are more likely to be scattered far away from the site
of their formation.  Our simulations show that both the close-in or farther out
planets should have lower mass than the planets with moderate
semi-major axes (Figs.~\ref{m_vs_a}, \ref{newmass_m_a}).  
This trend can be verified in future observations
using adaptive optics to detect and image giant planets further out
($40$--$100$ AU) from the central star \citep{2007ApJ...670.1367L}.
We find that a few percent of the simulated population have very high
semi-major axes in the final stable configuration (e.g.,
Fig.~\ref{a_vs_e}).  Such giant planets are extremely unlikely to be
created {\em in situ\/}, since the timescale for planet formation
greatly exceeds the age of the star \citep{2004MNRAS.347..613V}.
Additionally, there is simply insufficient disk mass to form a giant
planet at such large orbital distances
\citep{2002ApJ...581..666K,2004ApJ...616..567I,2004ApJ...604..388I}.
Strong scattering between planets in multi-planet systems can be a
natural mechanism to create such long-period planets
($a>50\,\rm{AU}$).  Our simulations suggest that this population of
high semi-major axis planets will have high eccentricities and
inclinations (Fig.~\ref{a_vs_e}).  Future planet searches using
astrometry or direct detection can test these predictions.  

We have also presented a preliminary study of the effects of a residual gas disk on 
planetary dynamics
(\S\ref{gas_planet}).  We compare the importance of dynamics for $9$  
different disk models with different disk surface
densities, keeping the initial orbital properties of the embedded
planets the same for all cases.  
We identify important timescales for the dynamical evolution of these 
systems.  In particular, we characterize the transitional stage of the 
dynamical evolution from the stable, eccentricity damped phase, where the 
planets are embedded in a massive disk to the unstable free eccentricity evolution 
stage following disk depletion.  
We show that it is possible to understand the overall evolution 
after planets are fully formed as an interplay between four 
different timescales, namely: the viscous timescale ($t_{vis}$) of the disk, 
the photo-evaporation timescale ($t_{photo}$) of the disk, 
the dynamical instability timescale $t_{dyn}$ for the 
planetary orbits, and the eccentricity damping timescale ($t_{damp}$) for the planetary orbits 
in a disk (\S\ref{gas_results}).  Our study clearly shows that 
planets will remain stable on nearly circular 
orbits while a sufficient amount of gas remains present in the disk, while with 
the same initial orbits without gas the system would become unstable.  The 
boundary between these two different phases can be characterized by $t_{damp}$, 
 $t_{vis}$, and $t_{photo}$.  The unstable phase starts when gas mass is sufficiently 
depleted so that $t_{damp}>t_{photo/vis}$.  We find that the transition within a 
disk from a small to a large $t_{damp}$ can be fairly quick once 
$t_{photo}<t_{vis}$ (Fig.~\ref{instability_growth}).  After photo-evaporation 
takes over the disk evolution, the system undergoes a quick transition.  
Until the critical mass for photo-evaporation 
is reached, planetary eccentricities remain close to 
zero independent of the disk mass and previous dynamical history.  Then, once the 
critical density is reached, the system behaves as if it had started with near-circular 
initial planetary orbits in a gas-free environment.  Thus, apart from 
highlighting the relative importance of these 
timescales for the evolution of planetary systems, our results justify typical 
initial conditions used in most studies of gas-free multi-planet systems, 
including our own (\S\ref{gas_free}).  
The initial properties for a gas-free system would be the orbital properties 
of the system as found at the boundary where $t_{damp}>t_{photo/vis}$, in this context.  

This work was supported by NSF Grant AST--0507727
at Northwestern University. Support for E.B.F.\ was provided by NASA
through Hubble Fellowship Grant HST-HF-01195.01A awarded by the Space
Telescope Science Institute, which is operated by the Association of
Universities for Research in Astronomy, Inc., for NASA, under contract
NAS 5-26555.

\appendix
\section{Sample Size Needed to Accurately Characterize the Eccentricity Distribution}
\label{size_statistic}

Due to the chaotic nature of n-body integrations and finite precision
of computer arithmetic, the long-term integration of an exact system
is impossible.  Instead, we must perform ensembles of n-body
integrations and interpret the results in terms of the statistical
properties of the outcomes of a set of similar n-body
systems.  Given that any study is based on a finite sample size, it is
important to recognize the limitations on the precision of various
statistics due to the limited number of integrations.  In the context
of this paper (and similar works) many simulations of similar
planetary systems are performed to generate a random sample of
outcomes, each of which can be compared to the observable properties,
such as the eccentricity and the semi-major axis of extrasolar
planetary systems.  Here we present an analysis of precision of
several statistics describing the eccentricity distribution as a
function of the number of n-body integrations performed.  Formally, our
estimated precision is applicable only to our specific choice for the
distribution of initial conditions.  Since this and previous studies
have shown that many variations of the planet-planet scattering model
result in similar eccentricity distributions, we expect that our
results can be applied to many similar studies to estimate the
accuracy of various statistical properties.  Our results should also
give a quantitative way to decide the required number of
simulations to estimate various statistical properties to a given
precision, and we expect this will be of use to other researchers
when formulating research plans.  Since the eccentricities of the
planetary systems are one of the most interesting properties, we focus
our attention on statistics describing the distribution of final
eccentricities.  However, the basic idea can be applied to any
statistic describing the masses or orbital properties of the simulated planetary
systems.

We will estimate the precision of several interesting statistics (the
mean, the standard deviation, the 5th ($P5$) and the 95th ($P95$)
percentiles of the distribution of final eccentricities).  We estimate
the ``true''  value for the underlying population  based on estimates
obtained making  use of our  full sample of $N=1515$  simulations (see
\S\ref{gas_free}).  We then estimate the same statistics based on $m$ subsets of
the full sample, where each subset  is a random sample of $n$ systems.
Next,  we  compare  the   statistic  estimated  from  each  subset  of
simulations  to  the  statistic  based  on the  full  population.   We
systematically change  the sample size $n$  for each subset,  so as to
explore  how precisely we  can estimate  a given  statistic  as a
function of the sample size.  

\begin{figure}
\begin{center}
\plottwo{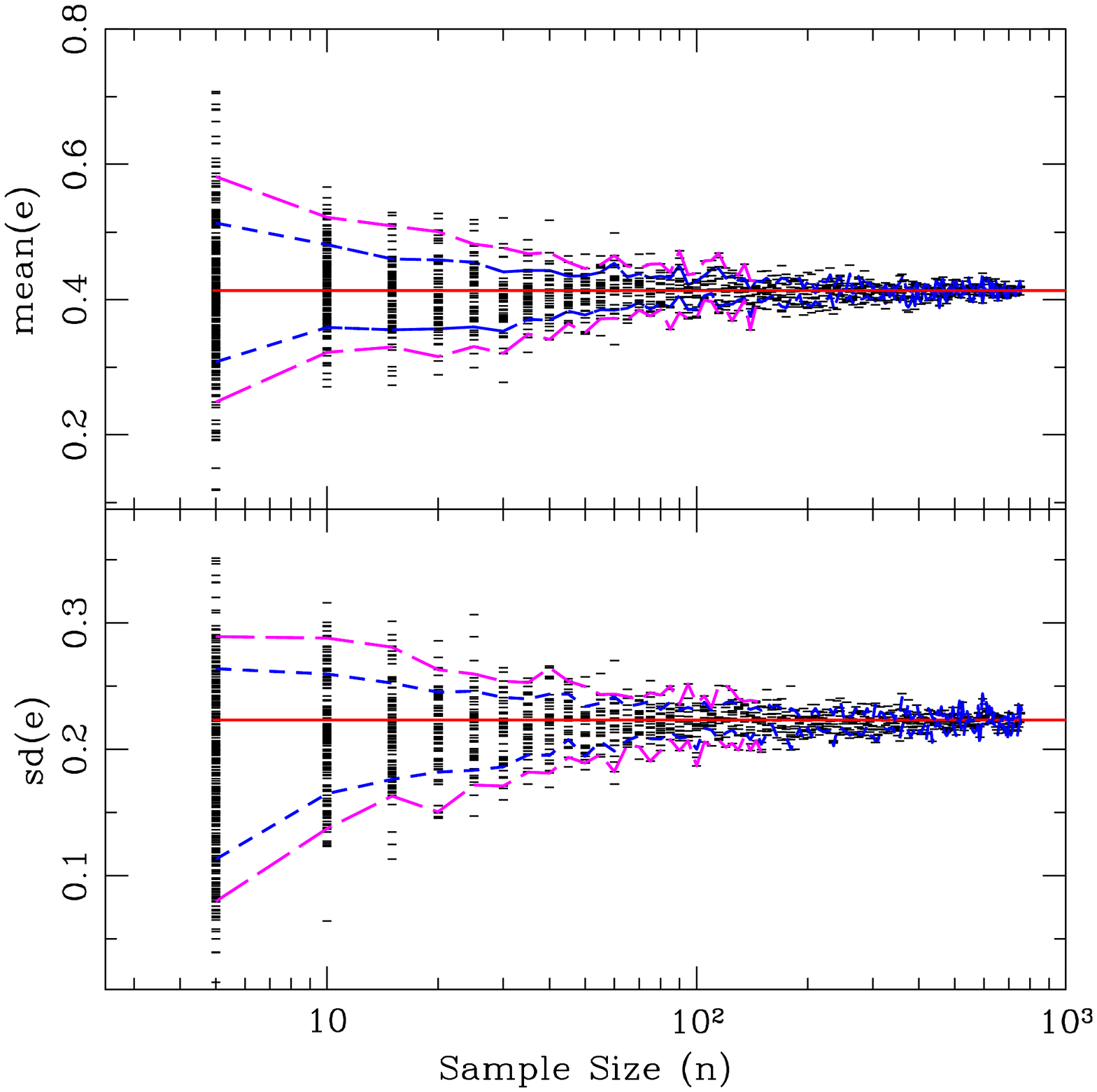}{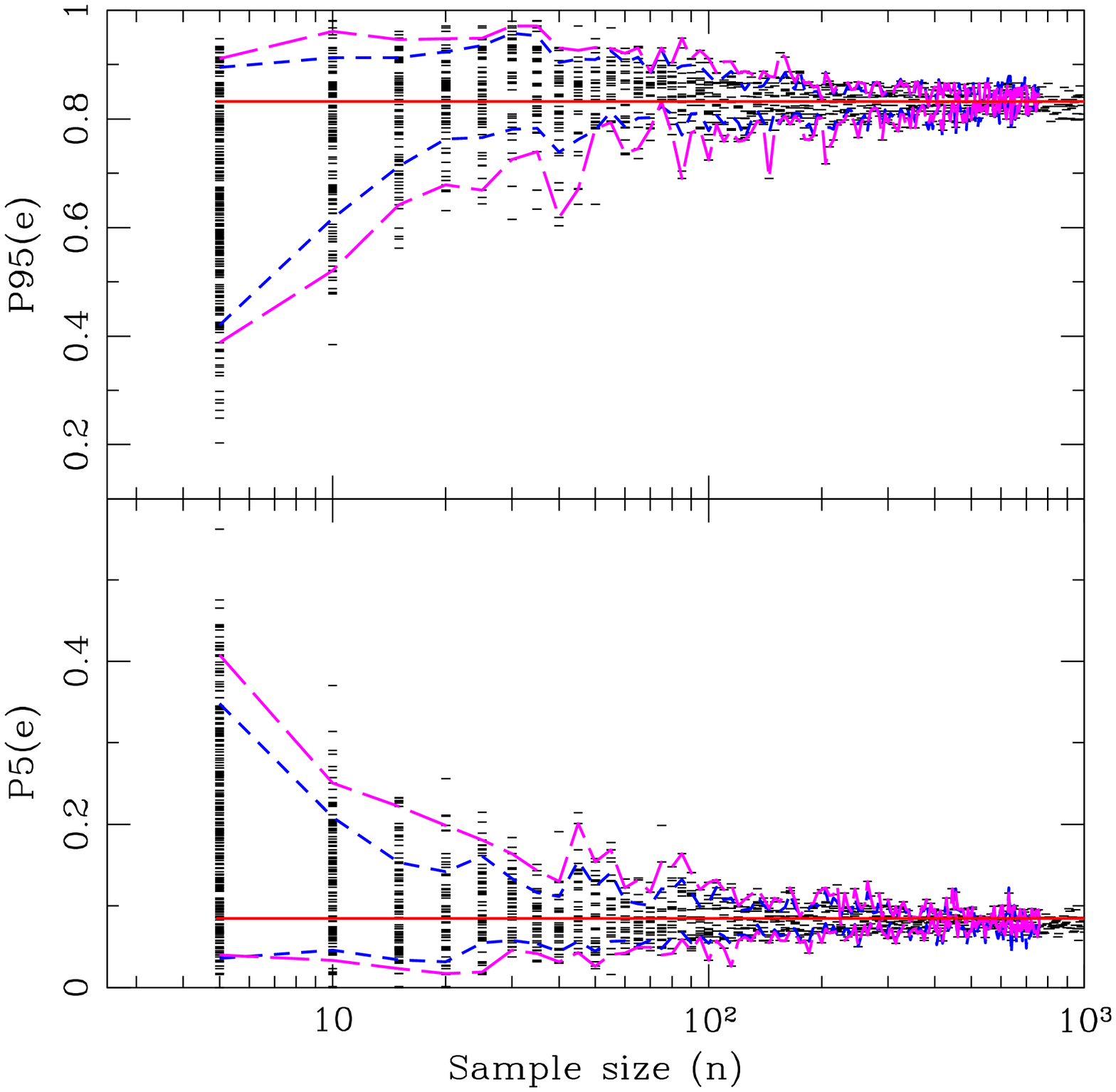}
\caption{{\em Top left panel:} Each dash shows the mean
eccentricities estimated from randomly chosen samples of sizes $n$,
specified by the x-axis.  The solid (red) line shows the mean eccentricity
of the full sample.  The short dashed (blue) lines show the
standard deviation in the estimates of the mean eccentricity with
respect to the full population mean for each sample size.  The long dashed (magenta) lines
show the 5th and the 95th percentiles of the estimates of the mean
eccentricities.  {\em Bottom left panel:} The same for the estimates of
the standard deviation for the eccentricities.  {\em Top right panel:} The same for
the estimates of the 95th percentiles for the distribution of eccentricities.
 {\em Bottom right panel:} The same for
the estimates of the 5th percentiles for the distribution of eccentricities.
\label{stat_scatter}
}
\end{center}
\end{figure}

\begin{figure}
\begin{center}
\plotone{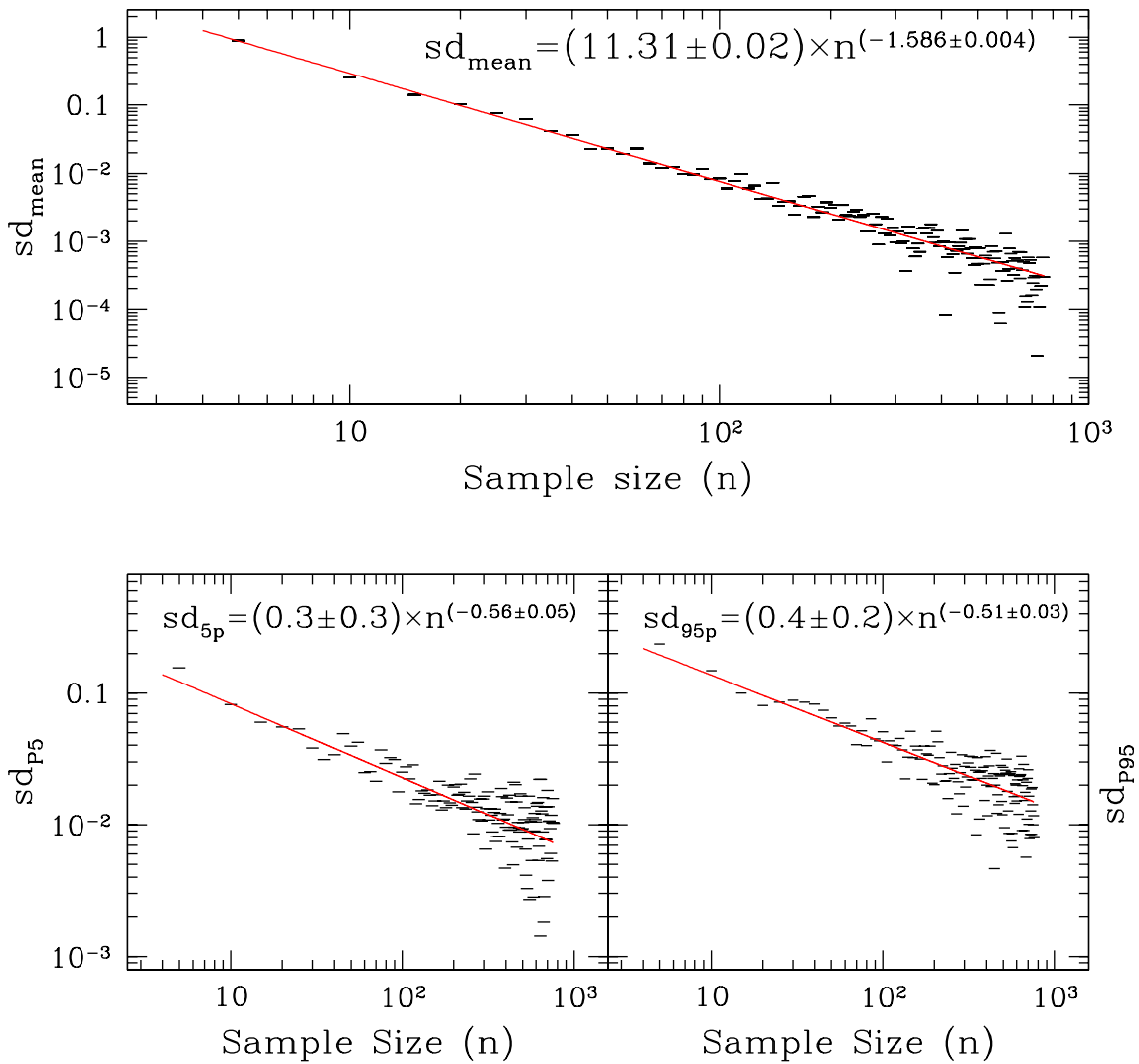}
\caption{Standard deviations in the estimates of the mean (top panel),
the 5th percentile (bottom left panel), and 
the 95th percentile (bottom right panel) for the eccentricity distribution
based on a randomly chosen sample of size $n$.  The solid (red)
lines show the empirical fit for the standard deviation of the
estimation of the statistic of interest as a function of the sample size.  The empirical
formulae for the fitting lines are also shown at the top of each
panel.  Note that the fits are only valid for samples of size greater
than $5$.  Moreover, the standard deviation in the estimates for
very large $n$ (and hence very small standard deviation) is limited by
the precision of our main simulated data (and is the reason for the
saturation seen in the standard deviation values near $n=1000$).
\label{stat_fit}
}
\end{center}
\end{figure}

In Fig. \ref{stat_scatter}, we show each
estimate  for a given  statistic as  a single  tick mark.   Each panel
presents  results  for  a  different  statistic:  mean  (left, top),
standard deviation  (left, bottom), $P95$ (right, top),
and $P5$ (right, bottom) of the eccentricity distribution.
For each of the above statistics,  we show the mean plus and minus the
standard deviation  (blue short  dashed curves) and  the 5th  and 95th
percentiles  (magenta long-dashed curves).   Hence, the  upper magenta
curve  of the  upper  left panel  shows  the $P95$ for  the
estimate  of  the mean  eccentricity  based  on  a sample  of  $m=N/n$
estimates of the mean eccentricity each using a sample size of $n$.
We find that for $n=50$ the standard error in estimating the
population mean from the sample can have a standard error of $37\%$,
whereas, for $n=100$ the precision improves to $5\%$
(Fig.~\ref{stat_scatter}).  In estimating the standard deviation of
the eccentricity from the small subsamples, we find that with $n=100$
the standard error in the estimation of the standard deviation of the
eccentricity distribution is $\sim 7\%$ (Fig.~\ref{stat_scatter}).
Although estimates for first two moments of the eccentricity
distribution can come within $10\%$ of the population moments based on
only $n\geqslant 100$, a much larger sample size is required to
characterize the tails of the eccentricity distribution accurately.
For example, the errors in estimating the $P5$ and
the $P95$ of the underlying eccentricity
distribution using sample sizes of $n=100$ are $30\%$ and $7\%$,
respectively.  If the sample size is increased to $n\sim1000$, then
$P5$ and $P95$ can be estimated within $4\%$ and $1\%$ error,
respectively,

Both previous and future studies often generate predictions for the
eccentricity distribution of planetary systems based on various
theoretical models.  For the sake of comparing the precision with
which these studies estimated the predicted eccentricity distribution,
we have used the above result to obtain an empirical relation between
the standard deviation of the estimates of various summary statistics
describing the eccentricity distribution and the number of simulations
used to estimate the statistic (Fig.~\ref{stat_fit}).  We expect that
this relation will also be useful for planning future studies, where
researchers will want to make a deliberate choice regarding the number
of simulations and other simulation parameters such as length of
integration time, number of particles, and inclusion of additional
physics.  We find that the standard deviation in the deviation of the
estimated mean eccentricity from the population mean eccentricity
decreases as a power law of sample size $n$ with an index of
$-1.586\pm 0.004$.  The standard deviation estimating
$P5$ and $P95$ decreases less steeply with $n$; here the power law
indices are $-0.58\pm0.05$ and $0.51\pm 0.03$, respectively.  (These
empirical relations are valid only for $n\geqslant 5$.)  For example,
a study that uses $100$ simulations would typically estimate the mean
of the predicted eccentricity distribution to within
$\simeq0.008$.  However, a larger number of simulations
becomes increasingly important for estimating the tails of the
eccentricity distribution precisely.  For example, an ensemble of 100
simulations typically estimates the $P5$ or the $P95$ with a
precision of $\simeq0.025$ or $\simeq0.045$, respectively.

\section{Stability Timescale}
\label{timescale}
According to the core accretion model of planet formation, planets
form in a protoplanetary disk separated by a small number of Hill
radii away from each other
\citep{1998Icar..131..171K,2002ApJ...581..666K}.  Hence, it is very interesting 
to have a good and statistically reliable investigation of the stability 
timescales as well as the distributions of the timescales as a function of the 
planet-planet distances in multiples ($K$) of their mutual Hill radii.  A similar timescale 
study was also performed by \citet{1996Icar..119..261C}.  However, their 
study covers a very different range of planetary masses.  The large ensembles 
used by our study not only produce a better statistical characterization of these 
timescales as a function of $K$, they also enable us for the first time to show the 
actual nontrivial shapes of these distributions.  The actual distributions 
of these timescales for a given $K$ value are particularly interesting 
for anyone performing a similar study and trying to decide upon a 
reasonable initial planetary separation, since due to the broad range, 
the computational effort needed will be determined by the few unusually 
stable realizations rather than the more frequent ones where instability 
can grow orders of magnitude quicker.  For better comparison with MW02 
we put the planet closest to the
star at $5\,$AU and then determine the semi-major axes of the other
two planets as follows,
\begin{equation}
\label{planet spacing}
a_{i+1} = a_i + K R_{H,i,i+1},
\end{equation}  
where ${K}$ is the spacing measured in terms of $R_{H,i,i+1}$, the mutual Hill radius for the 
$i^{th}$ and $i+1^{th}$ planets, 
\begin{equation}
\label{Hill radius1}
R_{Hi,i+1} = \left(\frac{M_i + M_{i+1}}{3 M_\star}\right)^{1/3} \frac{a_i + a_{i+1}}{2}, 
\end{equation}
following their prescription.  
Here $M_i$ is the mass of the $i^{th}$ planet, $M_\star$ the mass of
the central star, and $a_i$ the semi-major axis of the $i^{th}$
planet.  Note we use a different definition of Hill radius from that in \S\ref{mass_dist} 
following MW02 for easier comparison.

We integrate a number of 3-planet systems with different initial
conditions: $1000$ for $K\leqslant4.3$, $500$ for $4.3 <
K\leqslant5.0$ and $200$ for $K\geqslant 5.0$.  
Apart from the large number of realizations, we use a more general 
distribution of the initial eccentricities and orbital inclinations as described 
in \S\ref{mass_dist}.  

\begin{figure}
\begin{center}
\plotone{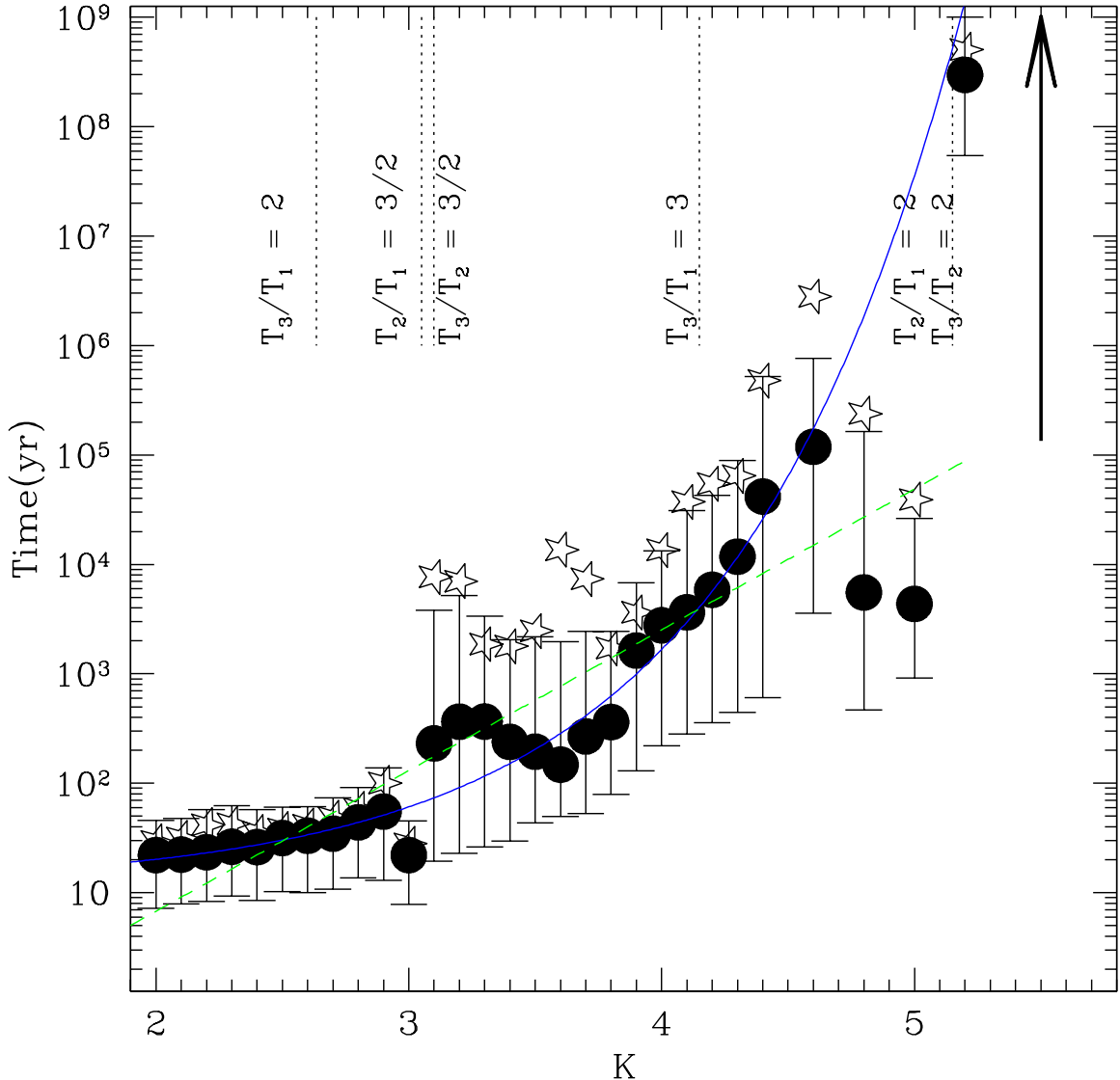}
\caption{Instability growth timescale as a function of spacing
parameter $K$.  The filled circles are the medians of the distributions
of stability timescales for the respective values of $K$.  The bars
show the $\pm 34\%$ range in the
timescale.  Note that they are not error bars, instead they represent the 
distribution of the stability timescales.  The medians are shown in the plot because the
distributions are skewed towards greater timescales.  
The open stars show the mean of the timescale
distribution.  Signatures of some major resonances can be noticed in the
sudden dips in timescale.  The arrow indicates that most systems with $K=5.5$ are stable for at least $10^9$ yr.  The starting point of the arrow represents the shortest instability timescale among our simulations.  The solid line (blue) and the dashed 
line (green) show the empirical best fit lines predicting the medians 
of the stability timescale distributions given the $K$ values 
(Eqs. ~\ref{exp_fit},\,\ref{lin_fit}).}
\label{t}   
\end{center}
\end{figure}

Fig.~\ref{t}
shows the results as a function of $K$.  The filled circles show the
median and the vertical bars above and below represent $\pm 34\%$
around the median.  Note that the vertical bars are not error
bars, but they are representative of the actual distributions of the
stability timescales.  We also show the mean of each distribution to
compare it with the median.  In each case the mean overestimates the
timescale and lies often outside the $34\%$ bars around the median.  
Our results are consistent with the findings of MW02 qualitatively.  We 
see similar trends near a MMR.  However, we find that a simple linear 
fit as was tried by MW02 does not work well.  A better empirical fit is given as 
follows.  
\begin{equation}
\label{exp_fit}
log_{10}t_m(K)=a+b\times exp(c\,K),
\end{equation}  
where $a$, $b$, and $c$ are constants (henceforth, called Fit-1).  The
best fit values for $a$, $b$, and $c$ (Table ~\ref{tvsk_fit}) can
predict the median timescales with fractional error less than $10\%$
away from MMR.  We also tried to find a simpler linear fit
$t_{m,linear}(K)$ for our data away from MMR, following MW02, writing
\begin{equation}
\label{lin_fit}
log_{10}t_{m,linear}(K)=a+b\,K,
\end{equation}     
where $a$ and $b$ are fitting parameters (henceforth, called Fit-2).
The best fit values for $a$ and $b$ are also given in Table
~\ref{tvsk_fit}.  We find that Fit-1 is much better than Fit-2.  In
particular, we find that linear fitting formula for instability
timescale as a function of initial spacing (as suggested by MW02) is
inaccurate by over three orders of magnitude for initial spacings such
that the planets are beyond the 2:1 MMR.  

\begin{figure}
\begin{center}
\plotone{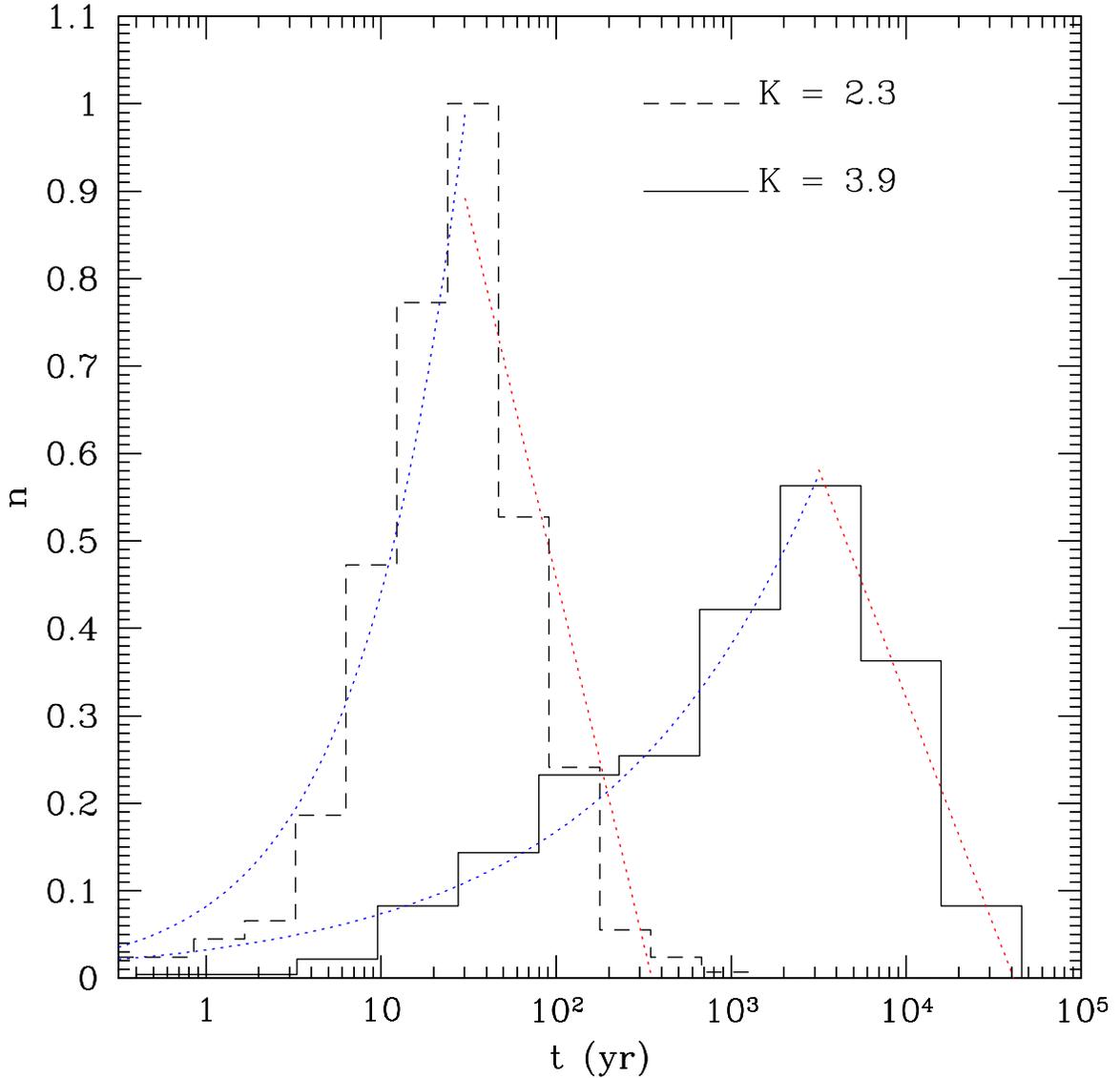}
\caption{Histograms for the timescale distributions at two different $K$
values both away from MMR.  Note that times are shown in log scale.
Each histogram shown here corresponds to $10^3$ runs for that $K$ value.
The number distributions are normalized such that $\sum_in_i\Delta
lt_i=1$, where, $\Delta lt_i$ is the bin-size in logarithm of time.
This normalization essentially makes the area under each histogram
normalized to $1$.  The solid histogram corresponds to $K=3.9$ and
the dashed histogram corresponds to $K=2.3$.  Both these $K$ values
are away from any major MMR.  The two histograms have essentially the
same shape.  The dotted (blue and red) curves show the analytical
fitting curves for timescale distributions at the left and the right
sides of the mode of the distributions.  For systems with stability
timescales less than the median of the distribution show an
exponential shape, whereas, those with timescales higher than the
median show a linear drop-off (see
Eqs. ~\ref{left_fit},\,\ref{right_fit} and Table ~\ref{tdist_fit}).  
\label{t_dist_noreso} }
\end{center}
\end{figure}

\begin{figure}
\begin{center}
\plotone{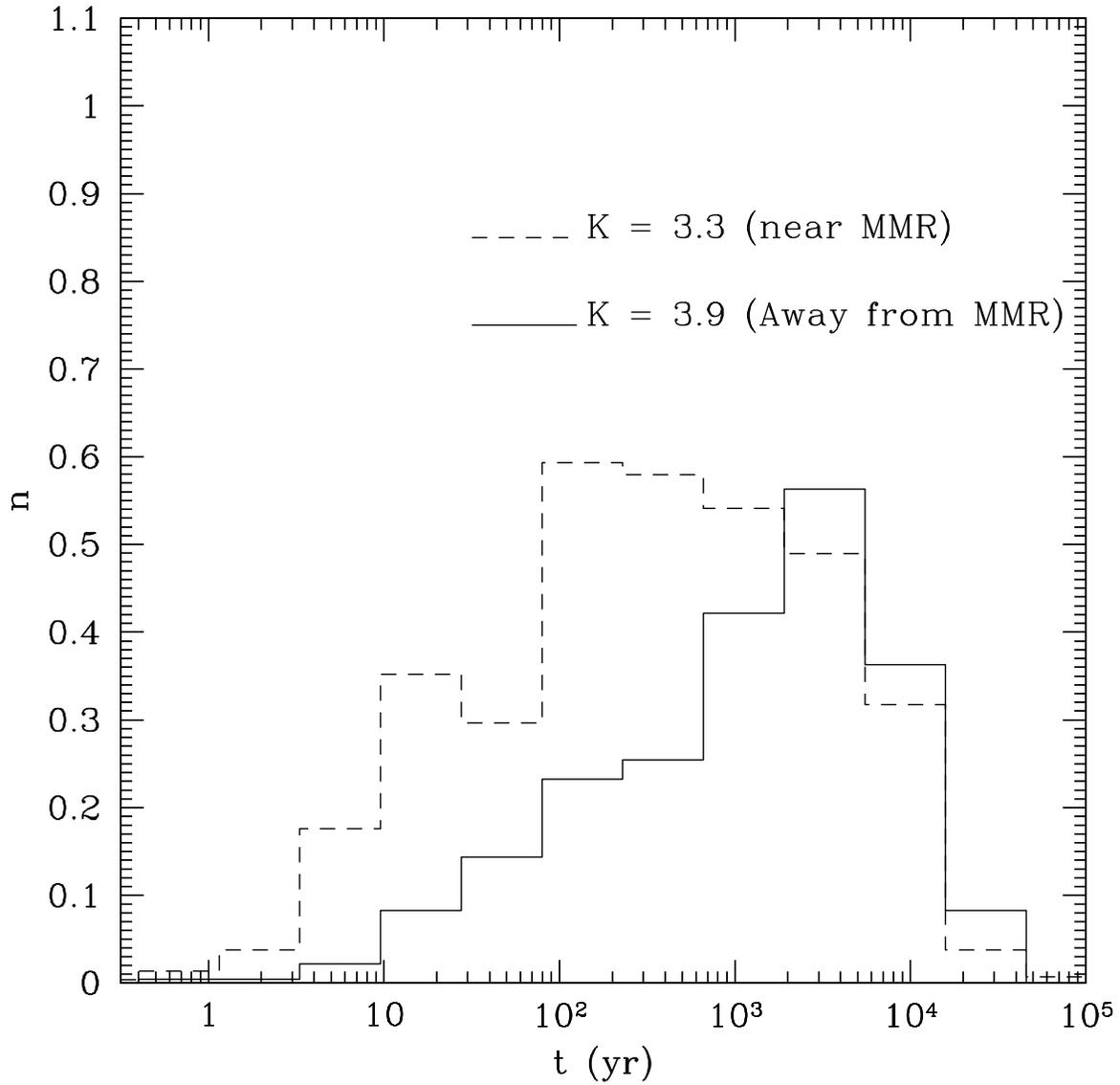}
\caption{Histograms for the timescale distributions near and away from an
MMR.  Each histogram corresponds to $10^3$ runs for that $K$ value.  We follow the 
same normalization scheme as mentioned earlier.  $K=3.3$ is near the $K$ value for a $3:2$ commensurability between the 
periods of the first and the second as well as the second and the third 
planetary orbits (dashed line).  
$K=3.9$ is away from MMR (solid line).  The distributions
near and away from MMR have somewhat similar shapes for times lower than 
the medians of the distributions.   However, for times higher than 
the medians the decay is not as 
sharp near a MMR as for systems far from a MMR.
}
\label{t_dist}   
\end{center}
\end{figure}

We also present the first study of the actual shapes of the timescale 
distributions.  In particular, in cases of
broad or skewed distributions, knowing only the median (or mean)
timescale can not provide a complete description of the distribution
of timescales to instability.  We find that the shapes of the
distributions of the timescales are essentially the same for any $K$
value away from a major MMR whereas near a major MMR the shape 
is very qualitatively different with a much slower decay above the median 
timescale (Figs.~\ref{t_dist_noreso}, \ref{t_dist}).  
Both
systems near and away from MMR show a similar exponential part in
the stability timescale distribution.  However, due to the MMR
configuration, some of the systems enjoy increased stability
manifested as a broader distribution to the higher time end.  
Note that the histograms of the timescale
distributions are normalized such that $\sum_i n_i\Delta lt_i=1$,
where $\Delta lt_i$ is the bin size in logarithm of time.  The
normalized number distribution for times lower than the median
timescale (henceforth denoted as $n_L$) has an exponential shape;
above the median timescale (henceforth denoted as $n_R$) the number
distribution has a linear decay for all $K$ away from major MMRs.  The
fitting formulae for $n_L$ and $n_R$ are given by
\begin{equation}
\label{left_fit}
n_L=N_L \exp\left[(\log_{10}t - \log_{10}t_m(K))/t_L\right],
\end{equation}
\begin{equation}
\label{right_fit}
n_R=N_R - t_R\,log_{10}t.
\end{equation}
Here, $N_{L}$ and $N_{R}$ are the normalization constants for the peak
amplitudes of the distributions, $t_m(K)$ is the median of the
timescale distribution as a function of $K$, $t_L$ and $t_R$ are
fitting constants characterizing the exponential index and the slope
of the two curves, respectively.  The best-fit values for $N_{L}$,
$N_{R}$, $t_{L}$, and $t_{R}$ are listed in Table ~\ref{tdist_fit}.
For a given $K$ value, the median timescale can be estimated using
Eq. ~\ref{exp_fit} and then using the median timescale the shapes of
the distributions can be obtained using Eqs. ~\ref{left_fit} and
~\ref{right_fit}.  

\begin{deluxetable}{ccccc}
\tablewidth{0pt}
\tablecaption{Fit: $t$ vs $K$ \label{tvsk_fit}}
\tablehead{
\colhead{ } & \colhead{a} & \colhead{b} & \colhead{c} & \colhead{Max Error(\%)}}
\startdata
Fit 1 & 1.07 & 0.03 & 1.10 & 10 \\
Fit 2 & -1.74 & 1.29 &  & 50 \\
\enddata
\tablenotetext{a}{The best fit values of the fitting parameters for the empirical 
fits for the median stability timescale of the systems as a function of their 
initial spacing parameter $K$.  }
\end{deluxetable}

\begin{deluxetable}{ccccc}
\tablewidth{0pt}
\tablecaption{Fit: Timescale distribution ($n_{L,R}$) \label{tdist_fit}}
\tablehead{
\colhead{$K$} & \colhead{$t_L$} & \colhead{$N_L$} & \colhead{$t_R$} & \colhead{$N_
R$}}
\startdata
2.0 & 0.719 & 0.714 & 1.728 & 3.716 \\
2.1 & 0.669 & 0.775 & 1.138 & 2.649 \\
2.2 & 0.582 & 0.886 & 0.778 & 1.971 \\
2.3 & 0.595 & 0.864 & 0.836 & 2.129 \\
2.4 & 0.663 & 0.867 & 1.193 & 2.816 \\
2.5 & 0.530 & 0.996 & 1.559 & 3.628 \\
2.6 & 0.482 & 1.153 & 1.286 & 3.065 \\ 
2.7 & 0.589 & 1.004 & 1.112 & 2.806 \\
2.8 & 0.582 & 1.002 & 1.167 & 3.01 \\
2.9 & 0.713 & 0.800 & 0.584 & 1.791 \\
3.0 & 0.624 & 1.715 & 1.412 & 3.141 \\
3.1 & 0.810 & 0.512 & 0.095 & 0.522 \\
3.2 & 2.343 & 0.206 & 0.082 & 0.484 \\
3.3 & 1.447 & 0.283 & 0.203 & 0.968 \\
3.4 & 1.051 & 0.462 & 0.188 & 0.889 \\ 
3.5 & 0.533 & 1.238 & 0.180 & 0.869 \\
3.6 & 0.395 & 3.700 & 0.201 & 0.929 \\
3.7 & 0.811 & 0.937 & 0.143 & 0.753 \\
3.8 & 0.744 & 1.085 & 0.245 & 1.140 \\
3.9 & 1.211 & 0.415 & 0.522 & 2.409 \\
4.0 & 1.488 & 0.353 & 0.349 & 1.811 \\
4.1 & 1.399 & 0.379 & 0.160 & 0.975 \\
4.3 & 1.684 & 0.346 & 0.204 & 1.292 \\
4.4 & 2.650 & 0.218 & 0.127 & 0.953 \\
4.8 & 1.066 & 4.910 & 0.102 & 0.710 \\
5.0 & 0.767 & 73.831 & 0.209 & 1.242 \\
\enddata
\tablenotetext{a}{Best fit values for the fitting parameters,
$N_L$, $t_L$, $N_R$, and $t_R$ predicting $n_L$, and $n_R$ for 
a given $K$ and the median of the stability timescale distribution $t_m(K)$.  
 }
\end{deluxetable}

\bibliography{biblio}

\begin{thebibliography}{76}
\expandafter\ifx\csname natexlab\endcsname\relax\def\natexlab#1{#1}\fi

\bibitem[{{Adams} \& {Laughlin}(2003)}]{2003Icar..163..290A}
{Adams}, F.~C. \& {Laughlin}, G. 2003, Icarus, 163, 290

\bibitem[{{Adams} \& {Laughlin}(2006{\natexlab{a}})}]{2006ApJ...649..992A}
---. 2006{\natexlab{a}}, \apj, 649, 992

\bibitem[{{Adams} \& {Laughlin}(2006{\natexlab{b}})}]{2006ApJ...649.1004A}
---. 2006{\natexlab{b}}, \apj, 649, 1004

\bibitem[{{Alexander} {et~al.}(2006){Alexander}, {Clarke}, \&
  {Pringle}}]{2006MNRAS.369..229A}
{Alexander}, R.~D., {Clarke}, C.~J., \& {Pringle}, J.~E. 2006, \mnras, 369, 229

\bibitem[{{Artymowicz}(1992)}]{1992PASP..104..769A}
{Artymowicz}, P. 1992, \pasp, 104, 769

\bibitem[{{Artymowicz}(1993)}]{1993ApJ...419..166A}
---. 1993, \apj, 419, 166

\bibitem[{{Black}(1997)}]{1997ApJ...490L.171B}
{Black}, D.~C. 1997, \apjl, 490, 171

\bibitem[{{Butler} {et~al.}(2006){Butler}, {Wright}, {Marcy}, {Fischer},
  {Vogt}, {Tinney}, {Jones}, {Carter}, {Johnson}, {McCarthy}, \&
  {Penny}}]{2006ApJ...646..505B}
{Butler}, R.~P., {Wright}, J.~T., {Marcy}, G.~W., {Fischer}, D.~A., {Vogt},
  S.~S., {Tinney}, C.~G., {Jones}, H.~R.~A., {Carter}, B.~D., {Johnson}, J.~A.,
  {McCarthy}, C., \& {Penny}, A.~J. 2006, \apj, 646, 505

\bibitem[{{Chambers}(1999)}]{1999MNRAS.304..793C}
{Chambers}, J.~E. 1999, \mnras, 304, 793

\bibitem[{{Chambers} {et~al.}(1996){Chambers}, {Wetherill}, \&
  {Boss}}]{1996Icar..119..261C}
{Chambers}, J.~E., {Wetherill}, G.~W., \& {Boss}, A.~P. 1996, Icarus, 119, 261

\bibitem[{{Clarke} {et~al.}(2001){Clarke}, {Gendrin}, \&
  {Sotomayor}}]{2001MNRAS.328..485C}
{Clarke}, C.~J., {Gendrin}, A., \& {Sotomayor}, M. 2001, \mnras, 328, 485

\bibitem[{{Cumming} {et~al.}(2008){Cumming}, {Butler}, {Marcy}, {Vogt},
  {Wright}, \& {Fischer}}]{2008arXiv0803.3357C}
{Cumming}, A., {Butler}, R.~P., {Marcy}, G.~W., {Vogt}, S.~S., {Wright}, J.~T.,
  \& {Fischer}, D.~A. 2008, arXiv:0803.3357

\bibitem[{{Duncan} {et~al.}(1998){Duncan}, {Levison}, \&
  {Lee}}]{1998AJ....116.2067D}
{Duncan}, M.~J., {Levison}, H.~F., \& {Lee}, M.~H. 1998, \aj, 116, 2067

\bibitem[{{Faber} {et~al.}(2005){Faber}, {Rasio}, \&
  {Willems}}]{2005Icar..175..248F}
{Faber}, J.~A., {Rasio}, F.~A., \& {Willems}, B. 2005, Icarus, 175, 248

\bibitem[{{Fabrycky} \& {Tremaine}(2007)}]{2007ApJ...669.1298F}
{Fabrycky}, D. \& {Tremaine}, S. 2007, \apj, 669, 1298

\bibitem[{{Ford} \& {Chiang}(2007)}]{2007ApJ...661..602F}
{Ford}, E.~B. \& {Chiang}, E.~I. 2007, \apj, 661, 602

\bibitem[{{Ford} {et~al.}(2001){Ford}, {Havlickova}, \&
  {Rasio}}]{2001Icar..150..303F}
{Ford}, E.~B., {Havlickova}, M., \& {Rasio}, F.~A. 2001, Icarus, 150, 303

\bibitem[{{Ford} {et~al.}(2000){Ford}, {Kozinsky}, \&
  {Rasio}}]{2000ApJ...535..385F}
{Ford}, E.~B., {Kozinsky}, B., \& {Rasio}, F.~A. 2000, \apj, 535, 385

\bibitem[{{Ford} \& {Rasio}(2006)}]{2006ApJ...638L..45F}
{Ford}, E.~B. \& {Rasio}, F.~A. 2006, \apjl, 638, L45

\bibitem[{{Ford} \& {Rasio}(2007)}]{2007astro.ph..3163F}
---. 2007, arXiv:astro-ph/0703163

\bibitem[{{Ford} {et~al.}(2003){Ford}, {Rasio}, \& {Yu}}]{2003ASPC..294..181F}
{Ford}, E.~B., {Rasio}, F.~A., \& {Yu}, K. 2003, in ASP Conf. Ser. 294:
  Scientific Frontiers in Research on Extrasolar Planets, ed. D.~{Deming} \&
  S.~{Seager}, 181--188

\bibitem[{{Gladman}(1993)}]{1993Icar..106..247G}
{Gladman}, B. 1993, Icarus, 106, 247

\bibitem[{{Goldreich} {et~al.}(2004){Goldreich}, {Lithwick}, \&
  {Sari}}]{2004ApJ...614..497G}
{Goldreich}, P., {Lithwick}, Y., \& {Sari}, R. 2004, \apj, 614, 497

\bibitem[{{Goldreich} \& {Sari}(2003)}]{2003ApJ...585.1024G}
{Goldreich}, P. \& {Sari}, R. 2003, \apj, 585, 1024

\bibitem[{{Goldreich} \& {Tremaine}(1980)}]{1980ApJ...241..425G}
{Goldreich}, P. \& {Tremaine}, S. 1980, \apj, 241, 425

\bibitem[{{Greenberg}(1974)}]{1974Icar...23...51G}
{Greenberg}, R. 1974, Icarus, 23, 51

\bibitem[{{Heap} {et~al.}(2000){Heap}, {Lindler}, {Lanz}, {Cornett}, {Hubeny},
  {Maran}, \& {Woodgate}}]{2000ApJ...539..435H}
{Heap}, S.~R., {Lindler}, D.~J., {Lanz}, T.~M., {Cornett}, R.~H., {Hubeny}, I.,
  {Maran}, S.~P., \& {Woodgate}, B. 2000, \apj, 539, 435

\bibitem[{{Hollenbach} {et~al.}(1994){Hollenbach}, {Johnstone}, {Lizano}, \&
  {Shu}}]{1994ApJ...428..654H}
{Hollenbach}, D., {Johnstone}, D., {Lizano}, S., \& {Shu}, F. 1994, \apj, 428,
  654

\bibitem[{{Holman} {et~al.}(1997){Holman}, {Touma}, \&
  {Tremaine}}]{1997Natur.386..254H}
{Holman}, M., {Touma}, J., \& {Tremaine}, S. 1997, \nat, 386, 254

\bibitem[{{Hut}(1980)}]{1980A&A....92..167H}
{Hut}, P. 1980, \aap, 92, 167

\bibitem[{{Ida} \& {Lin}(2004{\natexlab{a}})}]{2004ApJ...604..388I}
{Ida}, S. \& {Lin}, D.~N.~C. 2004{\natexlab{a}}, \apj, 604, 388

\bibitem[{{Ida} \& {Lin}(2004{\natexlab{b}})}]{2004ApJ...616..567I}
---. 2004{\natexlab{b}}, \apj, 616, 567

\bibitem[{{Ivanov} \& {Papaloizou}(2004)}]{2004MNRAS.353.1161I}
{Ivanov}, P.~B. \& {Papaloizou}, J.~C.~B. 2004, \mnras, 353, 1161

\bibitem[{{Juric} \& {Tremaine}(2007)}]{2007astro.ph..3160J}
{Juric}, M. \& {Tremaine}, S. 2007, arXiv:astro-ph/0703160

\bibitem[{{Kokubo} \& {Ida}(1998)}]{1998Icar..131..171K}
{Kokubo}, E. \& {Ida}, S. 1998, Icarus, 131, 171

\bibitem[{{Kokubo} \& {Ida}(2002)}]{2002ApJ...581..666K}
---. 2002, \apj, 581, 666

\bibitem[{{Lafreni{\`e}re} {et~al.}(2007){Lafreni{\`e}re}, {Doyon}, {Marois},
  {Nadeau}, {Oppenheimer}, {Roche}, {Rigaut}, {Graham}, {Jayawardhana},
  {Johnstone}, {Kalas}, {Macintosh}, \& {Racine}}]{2007ApJ...670.1367L}
{Lafreni{\`e}re}, D., {Doyon}, R., {Marois}, C., {Nadeau}, D., {Oppenheimer},
  B.~R., {Roche}, P.~F., {Rigaut}, F., {Graham}, J.~R., {Jayawardhana}, R.,
  {Johnstone}, D., {Kalas}, P.~G., {Macintosh}, B., \& {Racine}, R. 2007, \apj,
  670, 1367

\bibitem[{{Lee}(2004)}]{2004ApJ...611..517L}
{Lee}, M.~H. 2004, \apj, 611, 517

\bibitem[{{Lee} \& {Peale}(2002)}]{2002ApJ...567..596L}
{Lee}, M.~H. \& {Peale}, S.~J. 2002, \apj, 567, 596

\bibitem[{{Levison} \& {Morbidelli}(2007)}]{2007Icar..189..196L}
{Levison}, H.~F. \& {Morbidelli}, A. 2007, Icarus, 189, 196

\bibitem[{{Lin} \& {Ida}(1997)}]{1997ApJ...477..781L}
{Lin}, D.~N.~C. \& {Ida}, S. 1997, \apj, 477, 781

\bibitem[{{Lin} \& {Papaloizou}(1986)}]{1986ApJ...307..395L}
{Lin}, D.~N.~C. \& {Papaloizou}, J. 1986, \apj, 307, 395

\bibitem[{{Lissauer}(1995)}]{1995Icar..114..217L}
{Lissauer}, J.~J. 1995, Icarus, 114, 217

\bibitem[{{Marzari} \& {Weidenschilling}(2002)}]{2002Icar..156..570M}
{Marzari}, F. \& {Weidenschilling}, S.~J. 2002, Icarus, 156, 570

\bibitem[{{Matsumura} {et~al.}(2008){Matsumura}, {Thommes}, {Chatterjee}, \&
  {Rasio}}]{Matsumura08}
{Matsumura}, S., {Thommes}, E.~W., {Chatterjee}, S., \& {Rasio}, F.~A. 2008, in
  preparation

\bibitem[{{Mazeh} {et~al.}(1997){Mazeh}, {Krymolowski}, \&
  {Rosenfeld}}]{1997ApJ...477L.103M}
{Mazeh}, T., {Krymolowski}, Y., \& {Rosenfeld}, G. 1997, \apjl, 477, L103

\bibitem[{{Moorhead} \& {Adams}(2005)}]{2005Icar..178..517M}
{Moorhead}, A.~V. \& {Adams}, F.~C. 2005, Icarus, 178, 517

\bibitem[{{Moorhead} \& {Adams}(2008)}]{2008Icar..193..475M}
---. 2008, Icarus, 193, 475

\bibitem[{{Mouillet} {et~al.}(1997){Mouillet}, {Larwood}, {Papaloizou}, \&
  {Lagrange}}]{1997MNRAS.292..896M}
{Mouillet}, D., {Larwood}, J.~D., {Papaloizou}, J.~C.~B., \& {Lagrange}, A.~M.
  1997, \mnras, 292, 896

\bibitem[{{Murray} \& {Dermott}(2000)}]{2000ssd..book.....M}
{Murray}, C.~D. \& {Dermott}, S.~F. 2000, {Solar System Dynamics} (Cambridge
  University Press, 2000.)

\bibitem[{{Naef} {et~al.}(2005){Naef}, {Mayor}, {Beuzit}, {Perrier}, {Queloz},
  {Sivan}, \& {Udry}}]{2005ESASP.560..833N}
{Naef}, D., {Mayor}, M., {Beuzit}, J.-L., {Perrier}, C., {Queloz}, D., {Sivan},
  J.-P., \& {Udry}, S. 2005, in ESA Special Publication, Vol. 560, ESA Special
  Publication, ed. F.~{Favata} \& {et al.}, 833

\bibitem[{{Nagasawa} {et~al.}(2008){Nagasawa}, {Ida}, \&
  {Bessho}}]{2008arXiv0801.1368N}
{Nagasawa}, M., {Ida}, S., \& {Bessho}, T. 2008, arXiv:0801.1368

\bibitem[{{Narita} {et~al.}(2007{\natexlab{a}}){Narita}, {Enya}, {Sato},
  {Ohta}, {Winn}, {Suto}, {Taruya}, {Turner}, {Aoki}, {Tamura}, {Yamada}, \&
  {Yoshii}}]{2007astro.ph..2707N}
{Narita}, N., {Enya}, K., {Sato}, B., {Ohta}, Y., {Winn}, J.~N., {Suto}, Y.,
  {Taruya}, A., {Turner}, E.~L., {Aoki}, W., {Tamura}, M., {Yamada}, T., \&
  {Yoshii}, Y. 2007{\natexlab{a}}, arXiv:astro-ph/0702707

\bibitem[{{Narita} {et~al.}(2007{\natexlab{b}}){Narita}, {Sato}, {Ohshima}, \&
  {Winn}}]{2007arXiv0712.2569N}
{Narita}, N., {Sato}, B., {Ohshima}, O., \& {Winn}, J.~N. 2007{\natexlab{b}},
  arXiv:0712.2569

\bibitem[{{Ogilvie} \& {Lubow}(2003)}]{2003ApJ...587..398O}
{Ogilvie}, G.~I. \& {Lubow}, S.~H. 2003, \apj, 587, 398

\bibitem[{{Papaloizou} \& {Terquem}(2001)}]{2001MNRAS.325..221P}
{Papaloizou}, J.~C.~B. \& {Terquem}, C. 2001, \mnras, 325, 221

\bibitem[{{Rasio} \& {Ford}(1996)}]{1996Sci...274..954R}
{Rasio}, F.~A. \& {Ford}, E.~B. 1996, Science, 274, 954

\bibitem[{{S{\'a}ndor} \& {Kley}(2006)}]{2006A&A...451L..31S}
{S{\'a}ndor}, Z. \& {Kley}, W. 2006, \aap, 451, L31

\bibitem[{{S{\'a}ndor} {et~al.}(2007){S{\'a}ndor}, {Kley}, \&
  {Klagyivik}}]{2007A&A...472..981S}
{S{\'a}ndor}, Z., {Kley}, W., \& {Klagyivik}, P. 2007, \aap, 472, 981

\bibitem[{{Shakura} \& {Syunyaev}(1973)}]{1973A&A....24..337S}
{Shakura}, N.~I. \& {Syunyaev}, R.~A. 1973, \aap, 24, 337

\bibitem[{{Shu} {et~al.}(1993){Shu}, {Johnstone}, \&
  {Hollenbach}}]{1993Icar..106...92S}
{Shu}, F.~H., {Johnstone}, D., \& {Hollenbach}, D. 1993, Icarus, 106, 92

\bibitem[{{Simon} \& {Prato}(1995)}]{1995ApJ...450..824S}
{Simon}, M. \& {Prato}, L. 1995, \apj, 450, 824

\bibitem[{{Smith} \& {Terrile}(1984)}]{1984Sci...226.1421S}
{Smith}, B.~A. \& {Terrile}, R.~J. 1984, Science, 226, 1421

\bibitem[{{Takeda} \& {Rasio}(2005)}]{2005ApJ...627.1001T}
{Takeda}, G. \& {Rasio}, F.~A. 2005, \apj, 627, 1001

\bibitem[{{Terquem} \& {Papaloizou}(2002)}]{2002MNRAS.332L..39T}
{Terquem}, C. \& {Papaloizou}, J.~C.~B. 2002, \mnras, 332, L39

\bibitem[{{Thommes}(2005)}]{2005ApJ...626.1033T}
{Thommes}, E.~W. 2005, \apj, 626, 1033

\bibitem[{{Veras} \& {Armitage}(2004)}]{2004MNRAS.347..613V}
{Veras}, D. \& {Armitage}, P.~J. 2004, \mnras, 347, 613

\bibitem[{{Ward}(1993)}]{1993Icar..106..274W}
{Ward}, W.~R. 1993, Icarus, 106, 274

\bibitem[{{Ward}(1997)}]{1997ApJ...482L.211W}
---. 1997, \apjl, 482, L211

\bibitem[{{Weidenschilling} \& {Marzari}(1996)}]{1996Natur.384..619W}
{Weidenschilling}, S.~J. \& {Marzari}, F. 1996, \nat, 384, 619

\bibitem[{{Winn} {et~al.}(2006){Winn}, {Johnson}, {Marcy}, {Butler}, {Vogt},
  {Henry}, {Roussanova}, {Holman}, {Enya}, {Narita}, {Suto}, \&
  {Turner}}]{2006ApJ...653L..69W}
{Winn}, J.~N., {Johnson}, J.~A., {Marcy}, G.~W., {Butler}, R.~P., {Vogt},
  S.~S., {Henry}, G.~W., {Roussanova}, A., {Holman}, M.~J., {Enya}, K.,
  {Narita}, N., {Suto}, Y., \& {Turner}, E.~L. 2006, \apjl, 653, L69

\bibitem[{{Winn} {et~al.}(2005){Winn}, {Noyes}, {Holman}, {Charbonneau},
  {Ohta}, {Taruya}, {Suto}, {Narita}, {Turner}, {Johnson}, {Marcy}, {Butler},
  \& {Vogt}}]{2005ApJ...631.1215W}
{Winn}, J.~N., {Noyes}, R.~W., {Holman}, M.~J., {Charbonneau}, D., {Ohta}, Y.,
  {Taruya}, A., {Suto}, Y., {Narita}, N., {Turner}, E.~L., {Johnson}, J.~A.,
  {Marcy}, G.~W., {Butler}, R.~P., \& {Vogt}, S.~S. 2005, \apj, 631, 1215

\bibitem[{{Wolf} {et~al.}(2007){Wolf}, {Laughlin}, {Henry}, {Fischer}, {Marcy},
  {Butler}, \& {Vogt}}]{2007ApJ...667..549W}
{Wolf}, A.~S., {Laughlin}, G., {Henry}, G.~W., {Fischer}, D.~A., {Marcy}, G.,
  {Butler}, P., \& {Vogt}, S. 2007, \apj, 667, 549

\bibitem[{{Wright} {et~al.}(2007){Wright}, {Marcy}, {Fischer}, {Butler},
  {Vogt}, {Tinney}, {Jones}, {Carter}, {Johnson}, {McCarthy}, \&
  {Apps}}]{2007ApJ...657..533W}
{Wright}, J.~T., {Marcy}, G.~W., {Fischer}, D.~A., {Butler}, R.~P., {Vogt},
  S.~S., {Tinney}, C.~G., {Jones}, H.~R.~A., {Carter}, B.~D., {Johnson}, J.~A.,
  {McCarthy}, C., \& {Apps}, K. 2007, \apj, 657, 533

\bibitem[{{Wu} \& {Murray}(2003)}]{2003ApJ...589..605W}
{Wu}, Y. \& {Murray}, N. 2003, \apj, 589, 605

\bibitem[{{Zakamska} \& {Tremaine}(2004)}]{2004AJ....128..869Z}
{Zakamska}, N.~L. \& {Tremaine}, S. 2004, \aj, 128, 869

\end{thebibliography}

\end{document}